\documentclass[aps,prx,superscriptaddress,twocolumn,longbibliography,amsmath]{revtex4-2}

\usepackage{graphicx}% Include figure files

\usepackage{dcolumn}% Align table columns on decimal point
\usepackage{bm}% bold math
\usepackage{amssymb} % for \mathbb
\usepackage{graphicx}
\usepackage{xcolor}
\begin{document}

\title{Topological Insulator nano-SQUID: Flux-tunable platform for topological superconductivity}

\author{Ella Nikodem}
\thanks{These authors contributed equally to this work.}
\affiliation{Physics Institute II, University of Cologne, Z\"ulpicher Straße 77, 50937 K\"oln, Germany}

\author{Jakob Schluck}
\thanks{These authors contributed equally to this work.}
\affiliation{Physics Institute II, University of Cologne, Z\"ulpicher Straße 77, 50937 K\"oln, Germany}

\author{Henry~F.~Legg}
\affiliation{Department of Physics, University of Basel, Klingelbergstrasse 82, 4056 Basel, Switzerland}
\affiliation{SUPA, School of Physics and Astronomy, University of St Andrews,
North Haugh, St Andrews, KY16 9SS, United Kingdom}

\author{Max Geier}
\affiliation{Department of Physics, Massachusetts Institute of Technology, Cambridge MA 02139, USA}

\author{Micha{\l} Papaj}
\affiliation{Department of Physics, University of Houston, Houston, TX 77204, USA}

\author{Mahasweta~Bagchi}
\affiliation{Physics Institute II, University of Cologne, Z\"ulpicher Straße 77, 50937 K\"oln, Germany}

\author{Liang Fu}
\affiliation{Department of Physics, Massachusetts Institute of Technology, Cambridge MA 02139, USA}

\author{Yoichi Ando}
\email[]{ando@ph2.uni-koeln.de}
\affiliation{Physics Institute II, University of Cologne, Z\"ulpicher Straße 77, 50937 K\"oln, Germany}

\date{\today}

\begin{abstract}
Many efforts have been made in the past decade to realize topological superconductivity using superconducting proximity effect, but an ideal platform is still lacking. A 3D topological insulator (TI) is promising for this purpose due to the spin-momentum-locked surface state. Here we propose a novel yet simple TI platform which gives rise to a topological phase that is robust against disorder. It consists of a bulk-insulating rectangular TI nanowire laterally sandwiched by two superconductors. In this structure, the top and bottom surfaces individually work as SNS line junctions, forming a nanometer-scale columnar SQUID in which the nanowire cross-section defines the threading magnetic flux $\Phi$ in axial magnetic fields. We theoretically show that, when the two junctions are asymmetric, a robust topological phase occurs periodically for a wide range of $\Phi$, independently of the chemical potential. Our experiment found that a TI device of this structure indeed behaves as a columnar nano-SQUID where the supercurrent flows only through the top and bottom surfaces with vanishing bulk contribution. Furthermore, the top/bottom asymmetry can be tuned by a back gate, a key ingredient for the topological phase.

\end{abstract}


\maketitle

\section{Introduction}

The main feature of three-dimensional (3D) topological insulators (TIs) is the existence of gapless surface Dirac fermions that are protected by time-reversal symmetry \cite{Fu2007,Zhang2009,Ando2013}. These Dirac fermions are spin-non-degenerate with the spin direction dictated by the momentum; this spin-momentum locking is the key ingredient for realizing  ``spinless'' topological superconductivity that gives rise to Majorana zero-modes (MZMs) \cite{Fu2008}. The non-Abelian statistics obeyed by MZMs allows for topological quantum computing based on braiding operations \cite{Nayak2008}, so TIs have a high potential for such applications. However, TIs have been left behind in the main-stream efforts to realize topological superconductivity, largely due to the less developed technologies for TI device fabrications compared to semiconductor devices \cite{Breunig2021}. Unfortunately, despite significant efforts in the main-stream semiconductor platforms, the generation of MZMs is still elusive \cite{Flensberg2021, Prada2020}. This is mainly due to the sensitivity of the semiconductor platforms to Coulomb disorder \cite{Potter2011, Prada2020, Dassarma2023, Hess2023}, and a new platform that is inherently insensitive to chemical-potential fluctuations is strongly called for.

In this regard, Josephson junctions (JJs) are an interesting platform. A JJ consisting of a normal section in-between two superconducting (SC) electrodes is called an SNS junction. The properties of such a JJ are governed by the Andreev bound states (ABSs) that form in the normal metal as a result of multiple Andreev reflections. Supercurrent flows through an SNS junction because Cooper pairs can be transported via the ABSs. In the original proposal by Fu and Kane \cite{Fu2008} to realize MZMs in TIs, they considered 
an SNS line junction made on the TI surface. The ABSs in this junction become a 1D Majorana state that is $4\pi$-periodic as a function of the phase difference across the junction, $\varphi$, being topological for $\pi < \varphi < 3\pi$. 

Such a topological 1D Majorana state should in principle have a MZM when it has an end, but if the line junction ends in the middle of the TI surface, the MZM will hybridize with the gapless Dirac fermions in the rest of the surface; therefore, Fu and Kane proposed a phase-biased tri-junction to generate a MZM that is protected by a gap \cite{Fu2008}. A Josephson vortex in a TI line junction also hosts a MZM due to the naturally-occurring phase bias  \cite{Grosfeld2011, Potter2013}. These MZMs generated in a TI line junction have the important advantage that they are disorder-resilient \cite{Potter2011}, as their existence does not depend on the chemical potential at all, provided that the TI is bulk-insulating.
%Even though a protected MZM can be generated in this way, when the tri-junction is made on the top surface of a TI and the bottom surface is left normal (which is expected when the TI thickness is more than $\sim$20 nm \cite{Xu2014a, Xu2014b}), the gapless Dirac fermions at the bottom surface may compromise the parity of the device, causing problems for a Majorana qubit that encodes quantum information through fermion parity \cite{Alicea2012}. Similar problem exists when using the MZM appearing at the Josephson vortex core of a TI line junction \cite{Grosfeld2011, Potter2013}. Therefore, for topological quantum computing \cite{Nayak2008}, gapping out the whole surface is important \cite{Wu2024}. If this is achieved, the MZMs generated in a junction by phase biasing is robust and disorder-resilient \cite{Potter2011}, as their existence does not depend on the chemical potential at all provided that the TI is bulk-insulating. 
%Furthermore, the phase bias can be achieved by flowing supercurrents across the junction, so that the MZMs can, in principle, be generated without magnetic fields. 
%Recently, the 1D Majorana state of the Fu-Kane theory \cite{Fu2008} was confirmed by directly measuring the bound-state dispersion as a function of $\varphi$ in a TI line junction and was shown to be robust against the change in the chemical potential \cite{Schluck2024}.

\begin{figure*}
\centering
\includegraphics[width=0.9\textwidth]{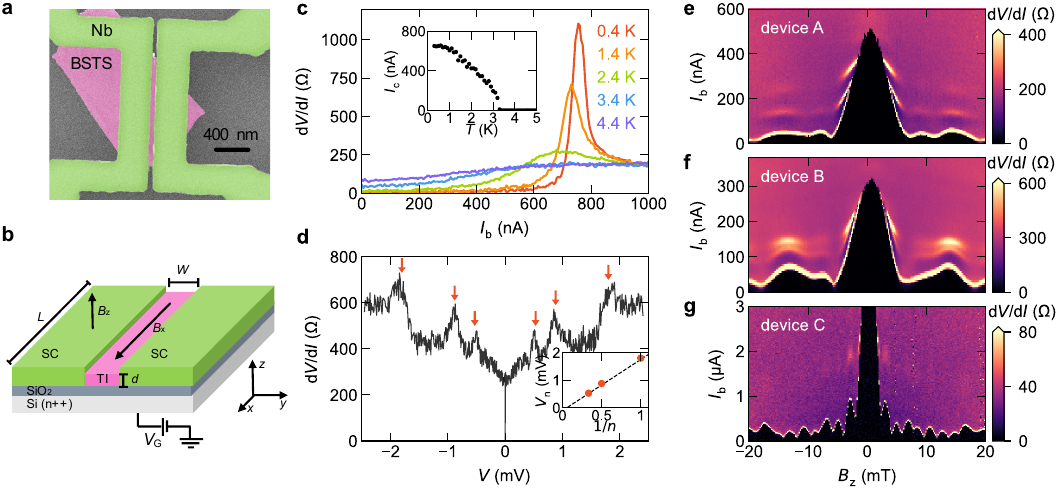}
\caption{{\bf Sandwich TINW Josephson junction.} (a) False-color SEM image of device A; an exfoliated TI flake is dry-etched to leave a NW (pink), and the etched area is subsequently filled with Nb (green), such that the TINW is side-contacted by superconducting Nb. There are remaining TI pieces that do not play any role. (b) Schematic of our device: Back-gated TINW is laterally sandwiched between two superconducting Nb electrodes to a Josephson junction. In- and out-of-plane magnetic fields can be applied. (c) $dV/dI$ vs $I_{\rm b}$ data in device A at various $T$ shown for zero back-gate voltage; inset shows the $T$-dependence of the critical current $I_c$ extracted from these data. We define $I_c$ as the bias current above which $dI/dV$ exceeded 50 $\Omega$.  
%The values of $I_\mathrm{c}$ were extracted from  d$I$/d$V$ vs $I_\mathrm{b}$ curves as currents at which d$I$/d$V$ reached a certain threshold. 
(d) Differential resistance $dV/dI$ vs junction voltage $V$ for device A in $B$ = 0 T. Red arrows mark the position of the peaks due to multiple Andreev reflection (MAR). We assign the index $n$ from 1 to 3 to the peaks ($n$ = 1 for the largest $V$). Inset shows a plot of the $V$ value for index $n$ (which we call $V_n$) vs $1/n$.  (e-g) Color mappings of $dV/dI$ as a function of $I_\mathrm{b}$ and $B_\mathrm{z}$ for devices A, B and C measured without any in-plane magnetic field at $T=30$\,mK, presenting Fraunhofer patterns. The critical current manifests itself as a bright orange line to border the black area corresponding to zero resistance. 
Irregular Fraunhofer features are attributed to an inhomogeneous supercurrent density along the junction as demonstrated in supplementary Fig. S5}. 

\label{fig:Fig1}
\end{figure*}

To exploit the benefits of the TI line junction in an alternative structure, we fabricate unconventional devices in which a rectangular nanowire, etch-fabricated from a bulk-insulating TI flake \cite{Feng2024}, is side-contacted by a superconductor on both left and right sides, creating a lateral sandwich SNS junction with a TI nanowire (TINW) as the N part (see Fig. 1b). In such a sandwich junction, if the TINW is bulk-insulating, the supercurrent flows only through the top and bottom surfaces (and a very small portion through the ends of the TINW) via the ABSs formed there. 
%The whole surface of this TINW is gapped, with the proximity-induced SC gap in the side surfaces and the mini-gap of the ABSs elsewhere \cite{Fu2008}. 
Since the axial magnetic field applied parallel to the TINW creates a magnetic flux $\Phi$ which controls the phase difference $\varphi$ at the top and bottom JJs, this structure can be viewed as a nanometer-scale columnar SQUID (superconducting quantum interference device) where the loop area corresponds to the TINW cross-section. A similar nano-SQUID was previously realized with a double-layer graphene heterostructure \cite{Indolese2020}. Our device has a bottom gate, which allows us to control the chemical potential in the TINW as well as to introduce asymmetry between the top and bottom junctions. A similar lateral sandwich junction was previously realized with a TI by using a diffusion process \cite{Bai2022}, but the device was not gate-tunable and it was not clear if the supercurrent flows only through the surface or also through the bulk.  

In our devices, we find that the critical current $I_c$ of the device exhibits pronounced oscillations as a function of $\Phi$ with a period $h/(2e)$ and, in certain situations, $I_c$ is almost completely suppressed when $\Phi=h/(4e)$ but recovers almost entirely for $\Phi=h/(2e)$. This is exactly the behavior expected for a nano-SQUID with completely surface-dominated supercurrent with negligible bulk contribution.  Theoretically, we first discuss a phenomenological SQUID model to explain the observed behavior, and then present numerical simulations based on a microscopic model of two line junctions made on a TINW surface, which support the SQUID model. 
Furthermore, we theoretically show that when the SQUID is asymmetric, the magnetic flux $\Phi$ causes only one of the two line junctions to go through the topological phase transition of the Fu-Kane type \cite{Fu2008} and the SQUID is topological for $(n-\frac{1}{2})\Phi_{0}^s < \Phi < (n+\frac{1}{2})\Phi_{0}^s$ with odd-integer $n$. This topological phase, which occurs for a wide parameter range, is robust against chemical-potential fluctuations due to disorder. The columnar nano-SQUID structure has an advantage over tri-junction or Josephson-vortex structures in that the MZMs appear at the ends of the TINW and are easier to access.
Note that there have also been related theoretical proposals to use the quasi-one-dimensional states of TINWs under superconducting proximity effect for generating MZMs \cite{Cook2011, Cook2012, deJuan2019, Legg2021, Heffels2023}; however, these rely on the realization of a well-defined subband structure, i.e., that electron states on the top and bottom surfaces are coherently connected, which is most likely not realized in our device.
%Given the robustness of the topological phase generated this way, our columnar nano-SQUID is a promising platform for Majorana physics.

\begin{figure*}[t]
\centering
\includegraphics[width=0.9\textwidth]{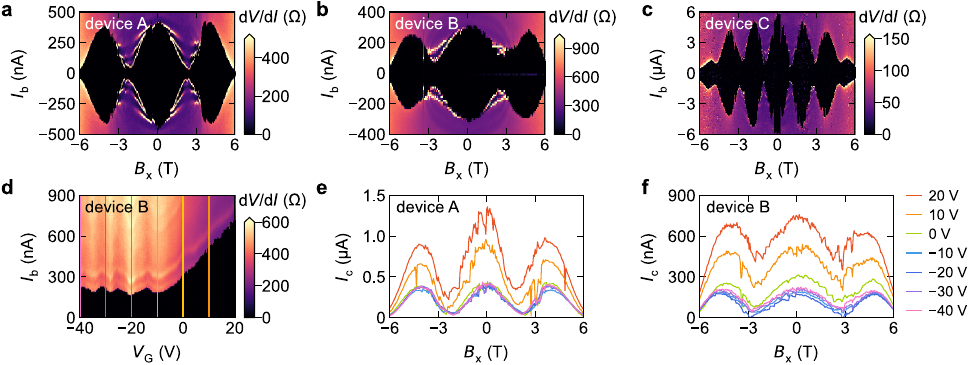}
\caption{{\bf Behaviour as a nano-SQUID.} (a-c) $dV/dI$ as a function of $I_\mathrm{b}$ and $B_\mathrm{x}$ obtained at $V_\mathrm{G}=0$\,V in devices A, B and C. The $B_z$-component due to misalignment is compensated to realize a purely in-plane magnetic field. There can be $\sim$1$^{\circ}$ misalignment towards the $y$ axis, but a small $B_y$ component does not affect the data nor the interpretation. The sweep direction of $I_\mathrm{b}$ is different for positive and negative currents, such that the sweep always starts from zero. (d) $dV/dI$ as a function of $I_\mathrm{b}$ and $V_\mathrm{G}$ for device A. Coloured vertical lines mark the positions where the $I_\mathrm{c}(B_x)$ behavior in panels (e) and (f) were measured. (e,f) $I_\mathrm{c}$ as a function of $B_x$ at various $V_\mathrm{G}$ values in devices A and B. All the data were acquired at $T=30$\,mK.}
\label{fig:Fig2}
\end{figure*}

\section{Columnar nano-SQUID}

{\bf Device characterization:} Our TINWs are based on bulk-insulating BiSbTeSe$_2$ \cite{Ren2011, Arakane2012} and are side-contacted by Nb electrodes.  A false-color scanning electron microscope (SEM) image of our device A is shown in Fig. \ref{fig:Fig1}a, while a device schematic is depicted in Fig. \ref{fig:Fig1}b along with the coordinate system. Note that the supercurrent flows {\it across} the TINW in our devices, while it flows {\it along} the wire in conventional devices \cite{Fisher2022, Roessler2023}. 
%We used a 6-1-1 T vector superconducting magnet for the precise alignment of the magnetic field along the TINW axis. 
We observed robust supercurrents at low temperature (Fig. 1c) and the plot of differential resistance $dV/dI$ vs junction voltage $V$ (Fig.~1d) presents signatures of multiple Andreev reflection (MAR). The energy positions of these MAR peaks (Fig.~1d inset) give the SC gap energy $\Delta_{\rm MAR} \approx 0.9$ meV, which is close to the SC gap of the Nb electrodes (1.0\,meV). From the $I$-$V$ characteristics, we can also extract \cite{Octavio1983, Flensberg1988} the averaged junction transparency $\tau_{\rm av}$; our analysis (see supplement) gives $\tau_{\rm av} \approx$ 0.76, which is reasonably high for a TI-JJ \cite{Ghatak2018}. These results demonstrate that the side-contact of a superconductor to a thin TI flake is possible and useful for the SC proximity effect. Similar side-contact technique is widely used for graphene \cite{Wang2013, Calado2015}, but it is new to TI flakes.

%Regarding the aspect of nanofabrication technology, our devices demonstrate that a side-contact of a superconductor to a thin TI flake ($\lesssim$20 nm thickness) can result in a high-transparency interface for the SC proximity effect. Such a side-contact technique is widely used for graphene \cite{Calado2015}, but it is new to TI flakes and will significantly widen the prospect of superconducting TI devices.

Importantly, our sandwich TINW junction forms a well-defined Josephson junction along the full length. This is evidenced by the Fraunhofer patterns shown in Figs.~1e-g for three devices. Here, the magnetic field $B_z$ is applied perpendicular to the junction plane. The $B_z$-dependence of $I_c$ presents clear nodes, which strongly suggests that the supercurrent is not due to superconducting shorts but the SC phase difference $\varphi$ winds along the junction (in the $x$-direction) in response to $B_z$. 
In all three devices, the $B_z$ values at the first $I_c$ minimum are essentially consistent with the expected $B_z$ values to create the SC flux quantum $\Phi_0^s = h/2e$ in the junction area, taking into account the flux-focusing effect \cite{Ghatak2018, Roessler2023, Rosenbach2021} (see supplement for details).
We note that the observed Fraunhofer pattern is not very regular in all three devices. This is understood as a result of inhomogeneity in the critical current along the junction \cite{Ghatak2018, Suominen2017, Beach2021}. This irregularity suggests that either the side of the etched TINW is not smooth or the induced pairing potential is nonuniform, resulting in locally varying $I_c$.
In the supplement, we show that the observed irregular Fraunhofer pattern can actually be reproduced with a position-dependent $I_c$ determined by a maximum-entropy fitting \cite{Ghatak2018}. The weak asymmetry between $\pm B_z$ is most likely an additional self-field effect \cite{Zhang2024}.
%The maximum in the Fraunhofer pattern should appear at $B_z$ = 0 even in the presence of an anomalous phase $\varphi_0$ \cite{Buzdin2008} that is expected for a TI Josephson junction in a strong $B_x$-field \cite{Assouline2019}. This is because the global phase offset will self-tune to maximize the critical current in a current-biased Josephson junction. Nevertheless, the anomalous phase $\varphi_0$ leads to an asymmetry in the Fraunhofer pattern across $B_z$ = 0 \cite{Assouline2019}, which is apparent in our data in Figs. 2A--E, suggesting that our sandwich TINW junction also develops an anomalous phase $\varphi_0$ in a large $B_x$. 

%\begin{table}[b]
%\centering
%\begin{tabular}{ c | c  c  c  c }
%Device  & $\Delta B_x^\mathrm{exp}$ (T)  & $S_\mathrm{geo}$ (nm$^2$)  & $S_\mathrm{eff}$ (nm$^2$) & $\Delta B_x^\mathrm{th}$ (T) \\
%\hline
%A & 5.0 & 780 & 797 & 5.2 \\
%B & 4.6 & 900 & 926 & 4.5 \\
%C & 1.9 & 1760 & 1838 & 2.2 \\
%\end{tabular}
%\caption{Summary of the experimental and theoretical $B_x$-periodicity to change the flux $\Phi$ in the TINW by $\Phi_0 = h/e$.}
%\label{table:Tab1}
%\end{table}

{\bf Critical-current oscillations in parallel magnetic fields:} 
To check for the SQUID behavior, we need to apply the magnetic field parallel to the TINW. 
This requires an accurate alignment, which was done by measuring $I_c$ as a function of small $B_\mathrm{z}$ in the presence of large nominal $B_x$ fields generated by our 6-1-1 T vector magnet (see supplement for details). By determining the necessary $B_z$ values to compensate for the misaligned $B_z$ component for large nominal $B_x$ values, we were able to completely cancel $B_z$ and obtain the $B_x$-dependence of the critical current in the absence of $B_z$. 
%As a side product of these calibrations, we obtained the $I_c(B_z)$ behavior in strong $B_x$-fields, which shows a signature of an anomalous phase $\varphi_0$ \cite{Buzdin2008, Assouline2019} (see supplement). This is an interesting topic of future research.

Figures 2a-c show the $I_c(B_x)$ behavior thus obtained in devices A, B, and C, that are slightly different in dimensions; device C was thicker and wider than others.
One can see that $I_c$ presents pronounced oscillations as a function of $B_x$, almost vanishing at the minima in devices A and C, but virtually recovers its zero-field value at the maxima. These strong oscillations of $I_c$ are completely different from the Fraunhofer pattern and are akin to the SQUID pattern. 
To understand the period $\Delta B_x$ of these $I_c$ oscillations, we evaluated the geometrical and effective cross-sectional area of a total of five TINW devices: The geometrical area $A_\mathrm{geo}$ is calculated with the actual width $W$ and the thickness $d$, while the effective area $A_{\rm eff}$ takes into account the penetration depth of the surface-state wavefunction; namely, we assumed that the effective electronic boundary of the nanowire is $\sim$2.5 nm deep from the outer surface \cite{Zhang2010crossover,Liu2010,Legg2022a}.

\begin{figure}[t]
\centering
\includegraphics[width=0.9\columnwidth]{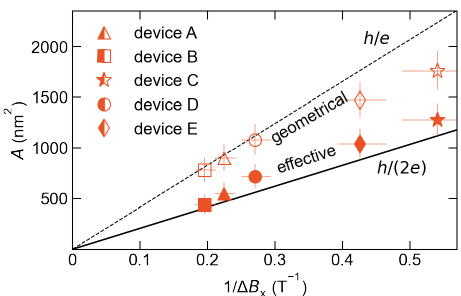}
\caption{{\bf Relation between the TINW cross-section and the $B_x$-period of the $I_c$ oscillations.} The period $\Delta B_x$, defined as the separation between the first minima for positive and negative $B_x$, is obtained from the $I_c$ oscillations in devices A-E (see supplement). The geometrical cross-section $A_\mathrm{geo}$ is simply $W$ times $d$. The effective cross-section $A_{\rm eff}$ takes into account the depth of the surface-state wavefunction ($\sim$2.5 nm)~\cite{Zhang2010crossover,Legg2022a}. $A_\mathrm{geo}$ and $A_\mathrm{eff}$ are plotted vs $1/\Delta B_x$ with open and filled symbols, respectively. Error bars represent the maximum estimated uncertainty; assuming uncertainties of 1 nm in $d$, 5 nm in $W$, and 100$-$200 mT (depending on the device) in the magnetic field for $I_c$ minima.} The solid and dashed lines are the expected relations when the period corresponds to the magnetic flux of $h/(2e)$ and $h/e$, respectively.
\label{fig:Fig3}
\end{figure}

In Fig.~3, $A_\mathrm{geo}$ and $A_\mathrm{eff}$ are plotted vs $1/\Delta B_x$, along with the lines showing the area corresponding to the flux change $\Delta \Phi=h/e$ and $\Delta \Phi=h/(2e)$. 
%We note that even the geometrical areas, that do not take into account the wavefunction extent, are consistently below $h/e$, suggesting that the oscillations are not of this frequency. 
One can easily see that the observed $\Delta B_x$ corresponds to the flux change of $h/(2e)$ if $A_{\rm eff}$ is taken. Thus, we conclude that the period of the $I_c$ oscillations is set by $h/(2e)$, which is expected for a SQUID. Note that the flux-focusing effect from the SC electrodes can be neglected in this geometry, because the Nb electrodes are almost fully penetrated by parallel magnetic fields. Note also that at the maximum $B_x$ field of this experiment, $\pm$6 T, the superconductivity of the Nb electrodes was about to disappear, making the $I_c$ lobes in Figs. 2a-c to be distorted and suppressed towards 6 T. The reasonable agreement of $A_{\rm eff}$ with the data suggests that possible fabrication damages to the surface are benign enough not to affect the surface-state penetration depth.

It is useful to mention that a similar SQUID pattern in the $I_c$ oscillations due to topological edge states has been reported for a 2D TI,  HgTe \cite{Hart2014}, and for a higher-order TI, Bi \cite{Murani2017}; in these cases, a planar junction structure resulted in a SQUID behavior due to the dominance of supercurrents through the 1D edge states. In our case, the SQUID behavior owes to the dominance of 2D surface states in a sandwich junction structure. Note that the necessary magnetic field to generate $h/(2e)$ flux in the SQUID-loop is significantly larger in our device.

One of the advantages of our geometry is that the chemical potential $\mu$ can be tuned by a back-gate voltage $V_{\rm G}$. The $V_{\rm G}$-dependence of $I_c$ shown in Fig.~2d for device B demonstrates that $\mu$ can be tuned across the Dirac point, which appears to be located at around $V_{\rm G} \approx -10$ V. The $I_c$ increases by a factor of 3 as $\mu$ moves away from the Dirac point into the $n$-type regime, while $I_c$ presents a saturation in the $p$-type regime. Such a behaviour is characteristic of gapless Dirac fermions and has been observed in planar JJs based on TIs \cite{Ghatak2018, Cho2013} and graphene \cite{Ben2016}; the saturation in the $p$-type regime was attributed to the formation of a $pn$-junction between the gated junction channel and the areas that are in direct contact with the SC electrodes. The weak oscillatory $I_c(V_{\rm G})$ behavior in the $p$-type regime could be due to a Fabry-Perot-like interference \cite{Calado2015, Ben2016}.

One can see in Figs.~2e-f that the pronounced $I_c$ oscillations as a function of $B_x$ are stable with respect to changes in $V_{\rm G}$, although an offset in $I_c$ grows with $V_{\rm G}$ in the $n$-type regime. Interestingly, device B is tuned with $V_{\rm G}\approx -20$ V to a regime where $I_c$ becomes zero at the minima. Such a disappearance of $I_c$ is possible only when there is no bulk contribution and the critical current at the top and bottom surfaces are equal. Hence, one can conclude that our TI nano-SQUID is surface-dominated.
This ``surface only" supercurent combined with its $V_{\rm G}$ dependence that is characteristic of gapless Dirac fermions are a clear manifestation of the TI nature.
Note that the formation of charge puddles in the bulk due to disorder is efficiently suppressed by the shielding effect of the superconducting electrodes \cite{Kaufhold2025}, which is an additional advantage of the nano-SQUID.

\begin{figure*}
\centering
\includegraphics[width=0.9\textwidth]{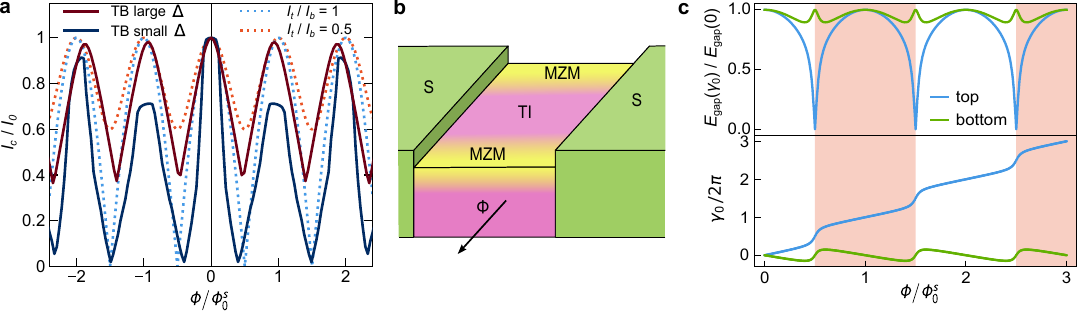}
\caption{{\bf Results of numerical simulations:} 
%(a) Examples of the CPR from the tight-binding model. Its maximum determines $I_c$. One can infer that $I_c$ is strongly suppressed at $\Phi$ close to $\Phi_0^s/2$ [$=h/(4e)$], but is almost recovered for $\Phi=\Phi_0^s=h/(2e)$. Importantly, for odd-integer flux $\Phi$, the ground state is realized for $\varphi = \pi$; the inversion of the CPR seen here is a result of this 0-$\pi$ transition. 
(a) The $I_c$ oscillations calculated for the simple phenomenological model (dotted lines) and the tight-binding model (solid line). The dominant period $\Delta \Phi=\Phi_0^s$ [$=h/(2e)$] appears in both models. In the phenomenological model Eq. (1), the ratio $I_t/I_b$ determines the size of the oscillations in $I_c$. The result of the tight-binding model for small $\Delta$ shows a superposition of oscillations with period $2\Phi_0^s$ [$=h/e$], which arises due to dependence of the normal state subbands on flux. This odd-even effect is not observed in our experiment, suggesting well-defined subbands is likely not formed. 
(b) Sketch: When the flux $\Phi$ is in the range $(n-\frac{1}{2})\Phi_{0}^s < \Phi < (n+\frac{1}{2})\Phi_{0}^s$ ($n$ odd), the columnar nano-SQUID hosts MZMs at the ends.
(c) Topological phase diagram, spectral gap of top- and bottom surfaces $E_{{\rm gap,t/b}}$, and gauge-invariant equilibrium phase difference $\gamma_{0, {\rm t/b}} = \varphi_0 - \frac{2\pi}{\Phi_0^s}\int {\bm A}{\rm d}{\bm l} = \varphi \pm \pi \Phi/\Phi_0^s$ at top- and bottom surfaces as function of magnetic flux $\Phi$ threading the nano-SQUID; calculated from Eq.~\eqref{eq:CPR}. Topological regions are shaded in red.}
\label{fig:Fig5}
\end{figure*}

{\bf Theoretical description:} 
Let us start with a simple phenomenological model to describe the columnar TI nano-SQUID. We assume two independent JJs on the top and bottom of the TINW and further assume that the side surfaces are proximitized by Nb to have a finite SC gap $\Delta$ and they connect the two JJs. This is obviously a SQUID with the loop-area set by the cross-section of the nanowire. If we consider only the first harmonic in the current-phase relation (CPR) for each JJ, the total current of this SQUID is written as
\begin{equation}
I(\varphi) \approx I_t \sin (\varphi-\pi \Phi/\Phi_0^s ) + I_b \sin(\varphi +\pi \Phi/\Phi_0^s ),
\label{eq:CPR}
\end{equation}
where $\varphi$ is the superconducting phase difference across the junctions and $I_t$ ($I_b$) is the maximum supercurrent through the top (bottom) surface.
The critical current can be obtained by finding the maximum as a function of $\varphi$, $I_c=\max_\varphi\{I(\varphi)\}$. The resulting $I_c$ as a function of $\Phi$ are shown in Fig.~4a for two ratios of $I_t/I_b$. As is usual for a SQUID, $I_c$ exhibits large oscillations with periodicity $h/2e$, and the depth of the $I_c$ minima at $\Phi=h/4e$ depends on the asymmetry, $ I_t / I_b$. We obtain $I_c \approx$ 0 at the minima in the symmetric case $ I_t \approx I_b$. Therefore, the observed $V_G$ dependence of the $I_c$ oscillations shown in Figs.~2e-f is understood to be essentially a result of the gate-induced change in $I_t$ and $I_b$.
For example, the data in Fig. 2f for $V_{\rm G}$ = 20 V and $-20$ V correspond to $I_t/I_b \simeq$ 0.5 and 1, respectively.

To support this phenomenological nano-SQUID model, we performed numerical simulations of our set-up by using a BHZ tight-binding model of a 3D TINW with finite cross-section and of infinite length, such that the momentum $k$ along the wire is a good quantum number (see Methods). We assume a spatial dependence of the pairing potential $\Delta(y)$ induced in the TINW surface, such that there is a phase difference of $\varphi$ across the top/bottom surface; namely, we assume the SC phase of $-\varphi/2$ for the left side surface and $+\varphi/2$ for the right. The ground state value of $\varphi$ is set by the flux $\Phi$, but it changes when a transverse current flows. This model effectively describes a SQUID with two JJs at the top and bottom of the TINW surface. The original TINW states are now broken into four parts, two at the sides with a SC gap and two at the top and bottom with ABSs. By diagonalizing the Bogoliubov-de Gennes Hamiltonian for this tight-binding model, we calculate the total energy $E$ as a function of $\varphi$, and its derivative gives the CPR, which allows us to calculate $I_c$ (see Methods and supplement).

Remarkably, this microscopic description reproduces the $I_c$ oscillations with a period of $h/2e$ and strong suppression near $h/(4e)$ as shown in Fig. 4a. 
This is because in our tight-binding calculations, between $\Phi = 0$ and $\Phi = \Phi_0^s$, the minimum of the energy-phase relation was found to change from $\varphi=0$ to $\varphi=\pi$; namely, there is a $0$--$\pi$ transition in the ground-state value of $\varphi$ due to the effect of $\Phi$ on the band structure
\footnote{The superconducting phase difference for our three-dimensional junction is defined gauge-invariantly by the mean of the local gauge-invariant phase difference $(1/h) \int_0^h dz \gamma(z)$ with $\gamma(z) = \phi - \int d{\bm l} e {\bm A}(y,z) / \hbar $. The mean gauge-invariant phase difference jumps (or quickly crosses over) from 0 to $\pi$ at $\Phi = \Phi_0^2/2$ flux bias.}. Importantly, at $\Phi \approx \Phi_0^s/2=h/(4e)$ where this $0$--$\pi$ transition happens, the total energy in our simulations is almost independent of $\varphi$ and we find $\partial E(\varphi )/\partial \varphi \approx 0$; this in turn results in a CPR that is extremely flat and $I_c$ becomes very small at $\Phi \approx h/(4e)$. 
Intuitively, upon current biasing, the flux $\Phi \approx \Phi_0^s/2$ forces the supercurrent to flow through the top and bottom surfaces in opposite directions; this tends to cancel the two contributions to the full current through the junction and leads to the minimum in $I_c$. This can be also viewed as a result of the CPR at the top and bottom to become out of phase upon the $0$--$\pi$ transition in $\varphi$.

We note that the $h/e$ periodicity of the original TINW states is visible when $\Delta$ is small, but it disappears when $\Delta$ is large. This is because a large SC gap in the side surfaces effectively decouples the top and bottom surfaces (see supplement for details).

\section{Topological Superconductivity}
 
We now show theoretically that the columnar TI nano-SQUID realizes a topological superconducting state over an extended flux range from $\Phi = (n-\frac{1}{2}) \Phi_0^s$ to $(n+\frac{1}{2}) \Phi_0^s$ around odd $n$ of flux quanta piercing the nanowire (see Fig. \ref{fig:Fig5}b for a sketch) and the only condition for the topological phase is a top/bottom asymmetry. Also, this topological phase is essentially insensitive to disorder.
This prediction is the most important result of this paper.

Recall that the spin-momentum locking in the TI surface states prohibits normal reflection at the NS interface for the perpendicular mode ($k_x$ = 0) and guarantees a pair of perfectly transmitted, non-degenerate ABSs (even- and odd-parity ABS) in the short-junction limit \cite{Fu2008}. They present $4\pi$ periodicity and cross zero-energy at $\varphi = (2m+1)\pi$ with integer $m$, where the fermion parity of the ground state alternates \cite{Beenakker2013Jan,Tkachov2013Aug}. When this $(2m+1)\pi$ crossing occurs in only one of the junctions in a columnar nano-SQUID, it guarantees the appearance of a 1D topological phase \cite{Kitaev2001Oct, Fu2008}. Now let us first consider the special case when an odd number of flux quanta pierces the nanowire, which leads to the superconducting phase to wind by an odd multiple of $2\pi$ in the nano-SQUID. The $2\pi$ relative phase difference (for example, phase zero at the top and $2\pi$ at the bottom) implies that the fermion parity on the top and bottom surfaces is opposite, making the SQUID topological. In this situation, the two ends of the columnar TI nano-SQUID become a boundary between the topological and trivial regions, which implies that MZMs should appear at the ends (Fig. 4b).

%Consider first the case of an odd number of flux quanta piercing the nanowire, which leads to the superconducting phase to wind by an odd multiple of $2\pi$ in the nano-SQUID. Because the TI surface states consist of a single, spin-momentum-locked Fermi surface that guarantees a perfectly transmitted channel in the short-junction limit, the $2\pi$ relative phase difference (for example, phase zero at the top and $2\pi$ at the bottom) implies that the fermion parity on the top and bottom surfaces is opposite. This is a consequence of the spectral winding of the non-degenerate, perfectly transparent Andreev bound state, whose zero-energy crossing at $\pi$ phase difference is protected by fermion-parity conservation \cite{Beenakker2013Jan,Tkachov2013Aug}. Changing the fermion parity of the $k=0$ mode in one of the junctions guarantees the appearance of a 1D topological phase \cite{Kitaev2001Oct, Fu2008}. In this situation, the two ends of the columnar TI nano-SQUID are a boundary between the topological and trivial regions, which implies that MZMs should appear at the ends.

In a closed system, topological phase transitions occur every half-integer flux quanta, $\Phi=(n + \frac{1}{2})\Phi_{0}^s,n\in\mathbb{Z}$.
At these points, time-reversal symmetry (TRS) is restored when all effects
of the magnetic field beyond the magnetic flux are neglected. Under this assumption, at $\Phi=(n + \frac{1}{2})\Phi_{0}^s$,  due to TRS, one of the Josephson junctions must have a phase difference $\pi$, while
the other has zero phase difference, which
corresponds to the topological phase transition point in the SQUID. The asymmetry
between top and bottom determines which surface turns topological: because
a $\pi$ phase difference switches the sign of the Josephson energy,
the junction with smaller Josephson energy acquires the $\pi$ phase
difference. 
%Related, the winding of the phase difference occurs only on the surface with a smaller Josephson energy, so that only this surface exhibits the topological phase transitions while the other remains trivial.
The topological region extends within a full flux quantum range
$(n-\frac{1}{2})\Phi_{0}^s < \Phi < (n+\frac{1}{2})\Phi_{0}^s$
around each odd-integer $n$ of flux quanta
threading the nano-SQUID, independent of the asymmetry between the top
and bottom surfaces. 

We verify the above picture by explicitly calculating the equilibrium
phase difference from the nano-SQUID model Eq.~(1) with
asymmetry $I_{{\rm t}}/I_{{\rm b}}=0.8$. The result shown
in Fig. \ref{fig:Fig5}c shows that the phase difference of the weaker top junction winds while the phase difference of the stronger bottom junction remains close to zero. 
Additionally, we show the spectral gap
of top and bottom surfaces which is given by the energy of the perfectly
transparent channel originating from the Dirac surface state, with
functional form $E_{{\rm gap,t/b}}(\varphi)\propto\cos(\varphi/2)$ \cite{Fu2008, Beenakker2013Jan}. The calculation confirms the spectral gap closure
at the top surface at $\Phi=\Phi_{0}^s/2$ while the gap on the other
surface remains open; this is the topological phase transition. 
Importantly, this calculation confirms the extended topological region for the flux bias, $(n-\frac{1}{2})\Phi_{0}^s < \Phi < (n+\frac{1}{2})\Phi_{0}^s$ with odd-integer $n$.
In future, it is useful to perform more detailed analysis to understand the exact role of the front and back surfaces of the TINW as well as the exact location of the MZMs in the topological phase.

%It is important to note that in our phenomenological model of the TI nano-SQUID, in the vicinity of fluxes $\Phi \approx  (2n+1)\Phi^s_0$ with integer $n$, the ground state is realized for $\varphi \approx \pi$. This means that each JJ should be at the topological phase transition, because a TI line junction becomes topological with odd parity for $\pi < \varphi < 3\pi$~\cite{Fu2008}. However, when the two JJs enter the topological phase at the same time, the total parity remains even and the TI nano-SQUID is topologically trivial. Nevertheless, when there is an asymmetry between the top and bottom junctions, their phases $\varphi_{\rm t}$ and $\varphi_{\rm b}$ will become different and their individual topological phase transition occurs at different values of $\Phi$. This means that there is a range of $\Phi$ where only one of the JJs is topological. In such a situation, the two ends of the columnar TI nano-SQUID are a boundary between the topological and trivial junctions, which means that MZMs should show up at the ends. 

%Although we do not have a means to detect the MZMs corresponding to the topological phase in the current version of our devices, the top/bottom asymmetry is tuned with the back gate, so the topological phase should have already been realized this experiment. It is an interesting future topic of research to theoretically elucidate the concrete condition to bring the TI nano-SQUID into the topological phase and to experimentally detect the expected MZMs.

\section{Robustness against disorder}

In experimental devices it is inevitable that disorder, e.g., in chemical potential and flux, will be present~\cite{Kaufhold2025}. We analyzed the stability of the topological phase by numerically simulating the energy gap to excited states (see supplement for details). We find that a gapped topological phase exists over a large region of the parameter space. In particular, the phase shows considerable stability against variations in chemical potential, persisting for the full range of the TI bulk band gap. The phase is also maintained across a sizable range of both the phase difference and the magnetic flux. This large region of phase space where topological superconductivity should be realised strongly suggests that the topological phase has a significant tolerance to the sources of disorder that can be expected in our devices. The largest energy excitation gaps are achieved for chemical potentials close to the charge neutrality point, when a flux quantum threads the nanowire, and for a phase difference of $\pi$ across the junction. The fact that this regime can actually be achieved in the current devices corroborates the promise of this platform to host robust topological superconductivity.

\section{Conclusion and outlook}

Our experiment demonstrates that the conceived columnar TI nano-SQUID can actually be fabricated with a new side-contacting technique and our device shows that the supercurrent is surface-dominated, which guarantees the prominent role of spin-momentum locking that dictates the appearance of perfectly-transmitted ABSs. Our theory shows that the occurrence of the topological phase transition in this nano-SQUID as a function of flux $\Phi$ is guaranteed by symmetry. Since the 1D topological superconductivity realized in a TI junction by phase-biasing is independent of chemical potential and the whole TI surface is gapped, our columnar nano-SQUID is a promising platform to host robust MZMs. An obvious next step is to experimentally confirm the emergence of the topological phase in the TI nano-SQUID, and a first experiment would be to detect a zero-bias conductance peak (ZBCP) due to MZMs with a tunnel probe. In this regard, it would be useful to theoretically investigate how a trivial ZBCP of non-MZM origin can appear in this platform. To get rid of trivial ZBCPs and to perform braiding to confirm the non-Abelian statistics of the MZMs, it will be important to realize a hard gap in this platform. Theoretical developments of concrete strategies to perform braiding in this promising platform would also be highly desirable.

\vspace{3mm}
\section*{Methods}

{\bf Experimental details:} 
The devices were fabricated on degenerately-doped Si substrates that were covered by a 290\,nm SiO$_2$ layer to enable the application of a global back-gate voltage. 
The TINWs are dry-etched with Ar plasma from mechanically-exfoliated thin flakes of the bulk-insulating TI material BiSbTeSe$_2$ \cite{Ren2011, Arakane2012}. They have a rectangular shape with width $W$ of 60 -- 70 nm and a length $L$ of a couple of micrometers. We mainly report three devices with slightly varying dimensions; the full dimensions of the TINWs used for this work are summarized in the supplement. The TINWs are side-contacted by Nb electrodes ($T_c\sim 7$\,K and $H_\mathrm{c2, in\mbox{-}plane} \gtrsim 6$\,T) on two sides and along the full length of the wire.  If not mentioned otherwise, all of the transport measurenents were performed at the base temperature ($\sim$30\,mK) of our dry dilution refrigerator. We used a 6-1-1 T vector superconducting magnet for the precise alignment of the magnetic field along the TINW axis, which was crucial for this work. The differential resistance $dV/dI$ was measured with a lock-in technique at 40 Hz by superimposing a small AC current to the DC bias current $I_{\rm b}$. 

{\bf Transport regime:} Fitting the data in the inset of Fig.~1c to $I_\mathrm{c}\sim\sqrt{T}\exp(-2\pi k_\mathrm{B}T/E_\mathrm{Th})$ gives the Thouless energy $E_\mathrm{Th} \approx$ 1.5\,meV. This is only slightly larger than the SC gap in Nb, implying that the junctions is in the intermediate to short junction regime \cite{Beenakker1992}. The SC coherence length $\xi$ in the top or bottom surface is estimated as $\xi_\mathrm{TI}=\sqrt{\hbar D/\Delta}=\sqrt{E_\mathrm{Th}L^2/\Delta} \approx$ 75\,nm ($L$ = 60 nm is the junction length). The $\xi$ on the side surface $\xi_{\rm side}$ should be much shorter, as the side surface is damaged by Ar etching. If we use $E_\mathrm{Th}=\hbar D/L^2$ with $D=v_\mathrm{F}l_\mathrm{e}/2$ for the estimate of the mean free path $l_\mathrm{e}$, we obtain $l_\mathrm{e}\approx30$\,nm for $v_\mathrm{F}=5.5\times 10^{5}$\,m/s \cite{Arakane2012}. Hence, we find $l_e < L \lesssim \xi_{\rm TI}$. 

{\bf Theory:} To treat the TINW, we use a model for the 3D TI material BSTS~\cite{Zhang2009, deJuan2019, Legg2021, Legg2022} with finite square cross-section of side length $2L+1$ in the Kwant package \cite{Groth2014}, such that the Bogoliubov-de Gennes Hamiltonian is given by
\small
\begin{equation}
\begin{aligned}
H_k & =\frac{1}{2} \sum_{\substack{n=-L \\
m=-L}}^{L, L} \mathbf{c}_{n, m, k}^{\dagger} \cdot\left\{M(k) \tau_z+A \sin (k) \tau_x \sigma_x-\mu \right\} \eta_z \mathbf{c}_{n, m, k} \\
+& \sum_{\substack{n=-L \\
m=-L}}^{L-1, L}\left\{\mathbf{c}_{n+1, m, k}^{\dagger} \cdot\left\{-B \tau_z+\frac{i A}{2 } \tau_x \sigma_y\right\} e^{i  \eta_z  \phi_m} \mathbf{c}_{n, m, k}\right\}\eta_z \\
+& \sum_{\substack{n=-L \\
m=-L}}^{L, L-1}\left\{\mathbf{c}_{n, m+1, k}^{\dagger} \cdot\left\{-B \tau_z+\frac{i A}{2} \tau_x\sigma_z\right\} \mathbf{c}_{n, m, k}\right\}\eta_z\\
+&\frac{1}{2}\sum_{\substack{n=-L \\
m=-L}}^{L, L} \mathbf{c}_{n, m, k}^{\dagger} \cdot  \Delta(y)\left\{\cos[\varphi(y)]\eta_x+\sin[\varphi(y)]\eta_y \right\}  \mathbf{c}_{n, m, k}\\
+&\text { H.c.,}\label{BHZ}
\end{aligned}
\end{equation}
\normalsize
where $M(k)=M_0-2B[\cos(k)-3]$ and the Pauli matrices $\tau_i$, $\sigma_i$, and $\eta_i$ act in orbital, spin, and particle-hole space, respectively. To model the nano-SQUID, we take the pairing potential to be finite on either side of the TINW, but zero for sites in the center
\begin{equation}
\Delta(y)=\begin{cases}
0 \;\;\;{\rm if}\;n\leq N\\
\Delta \;\;\;{\rm otherwise}
\end{cases}
\end{equation}
and the superconducting phase is given by
\begin{equation}
\varphi(y)=\begin{cases}
\varphi/2 \;\;\;{\rm if}\;n\geq0\\
-\varphi/2 \;\;\;{\rm if}\;n<0.
\end{cases}
\end{equation}
Finally, the Peierls term $\phi_m$ is set by the flux $\Phi$ such that,
\begin{equation}
\phi_m= \frac{4m\pi \Phi}{(2L+1)^2}.
\end{equation}
For simplicity we choose parameters $A=1$, $M=-1.5$, $B=0.5$, $L=4$, and $\mu=0.5$. We choose the pairing potentials $\Delta=0.02$, which is smaller than the subband gap, and $\Delta=0.25$, which is larger than the subband gap (see Fig.~\ref{fig:Fig5}). We set $N$=3 such that only the side sites in the cross-section have an induced pairing potential. It should also be noted that, due to the finite extent of the wavefunctions into the nanowire, the value of flux corresponding to the normal-state flux quantum $\Phi=\Phi_0^n=h/e$ is obtained from the normal-state band structure (see supplement).

From the tight-binding model in Eq.~\eqref{BHZ} we can obtain the energy dispersion $E_i(k)$, where $i$ is a band index, for a given value of flux $\Phi$ and SC phase difference $\varphi$. The full energy-phase relation is then obtained from the energy of the occupied states
\begin{equation}
E(\varphi)=\int^\Lambda_{-\Lambda}dk \sum_{E_i(k)<0} E_i(k),\label{enphase}
\end{equation}
which assumes zero temperature. This energy-phase relation is then used to obtain the CPR (shown in Fig. 4a) as its $\varphi$-derivative. Note that Eq.~\eqref{enphase} utilises a momentum cutoff $\Lambda$, which is set such that the CPR has converged.

\section*{Acknowledgments}
We thank Oliver Breunig for the help in Fraunhofer-pattern fittings. This project has received funding from the European Research Council (ERC) under the European Union’s Horizon 2020 research and innovation program (Grant Agreement No. 741121) and was also funded by the Deutsche Forschungsgemeinschaft (DFG, German Research Foundation) under Germany's Excellence Strategy - Cluster of Excellence Matter and Light for Quantum Computing (ML4Q) EXC 2004/1 - 390534769, as well as by the DFG under CRC 1238 - 277146847 (Subprojects A04 and B01). E.N. acknowledges support by the Studienstiftung des deutschen Volkes. The work at Massachusetts Institute of Technology was supported by a Simons Investigator Award from the Simons Foundation. 

{\bf Data availability:} The data that support the findings of this study are available at the online depository zenodo with the identifier {10.5281/zenodo.14331680} and Supplementary Information.

\bibliography{nanoSQUID}% Produces the bibliography via BibTeX.

\end{document}

%{\bf{\large Supplemental Material for ``Topological Insulator nano-SQUID: Flux-tunable platform for topological superconductivity''}}

\title{Supplemental Material for ``Topological Insulator nano-SQUID: Flux-tunable platform for topological superconductivity''}

\author{Ella Nikodem}
\thanks{These authors contributed equally to this work.}
\affiliation{Physics Institute II, University of Cologne, Z\"ulpicher Straße 77, 50937 K\"oln, Germany}

\author{Jakob Schluck}
\thanks{These authors contributed equally to this work.}
\affiliation{Physics Institute II, University of Cologne, Z\"ulpicher Straße 77, 50937 K\"oln, Germany}

\author{Henry~F.~Legg}
\affiliation{Department of Physics, University of Basel, Klingelbergstrasse 82, 4056 Basel, Switzerland}
\affiliation{SUPA, School of Physics and Astronomy, University of St Andrews,
North Haugh, St Andrews, KY16 9SS, United Kingdom}

\author{Max Geier}
\affiliation{Department of Physics, Massachusetts Institute of Technology, Cambridge MA 02139, USA}

\author{Micha{\l} Papaj}
\affiliation{Department of Physics, University of Houston, Houston, TX 77204, USA}

\author{Mahasweta~Bagchi}
\affiliation{Physics Institute II, University of Cologne, Z\"ulpicher Straße 77, 50937 K\"oln, Germany}

\author{Liang Fu}
\affiliation{Department of Physics, Massachusetts Institute of Technology, Cambridge MA 02139, USA}

\author{Yoichi Ando}
%\email[]{ando@ph2.uni-koeln.de}
\affiliation{Physics Institute II, University of Cologne, Z\"ulpicher Straße 77, 50937 K\"oln, Germany}
\vspace{-0.2cm}

\maketitle

%{\large{\bf Materials and Methods}}

\vspace{-0.5cm}
\subsection{Materials and device fabrication}

To synthesize bulk single crystals of BiSbTeSe$_2$, we applied the modified Bridgman method to high-purity (99.9999\%) Bi, Sb, Te, and Se following the recipe described in Ref.\,\cite{Ren2011}. Thin flakes with thickness $d\lesssim 20$\,nm were mechanically exfoliated from these bulk single crystals onto degenerately-doped Si wafers covered by 290\,nm of SiO$_2$, which serves as dielectric for back-gating. Suitable flakes were identified by optical microscopy. Using electron beam lithography (Raith PIONEER Two), we defined masks to dry-etch the BiSbTeSe$_2$ flakes into rectangular nanowires with the width $W$ of 60 -- 80\,nm and length $L$ of a few micrometers with Ar plasma. Precise dimensions of the nanowires presented in this work are shown in Table \ref{table:Tab1}; $W$ and $L$ were determined with a scanning electron microscope (SEM) and $d$ with an atomic force microscope (AFM). 

\addtolength{\tabcolsep}{5pt}    
\begin{table}[h]
\small
    \centering
    \begin{tabular}{ cl l l l}
    Device & $d$ (nm) & $W$ (nm) & $L$ ($\mu$m) \\ 
    \hline
    A & 15 & 60 & 1.5\\  
    B & 13 & 60  & 1.2\\  
    C & 22 & 80 & 2.4 \\
    D & 18 & 60 & 1.4 \\
    E & 21 & 70 & 1.9 \\
\end{tabular}
\caption{Nanowire dimensions for devices A--E.}
\label{table:Tab1}
\end{table}
\normalsize

\vspace{-0.5cm}
\begin{figure}[h]
	\centering
	\includegraphics[width=0.72\textwidth]{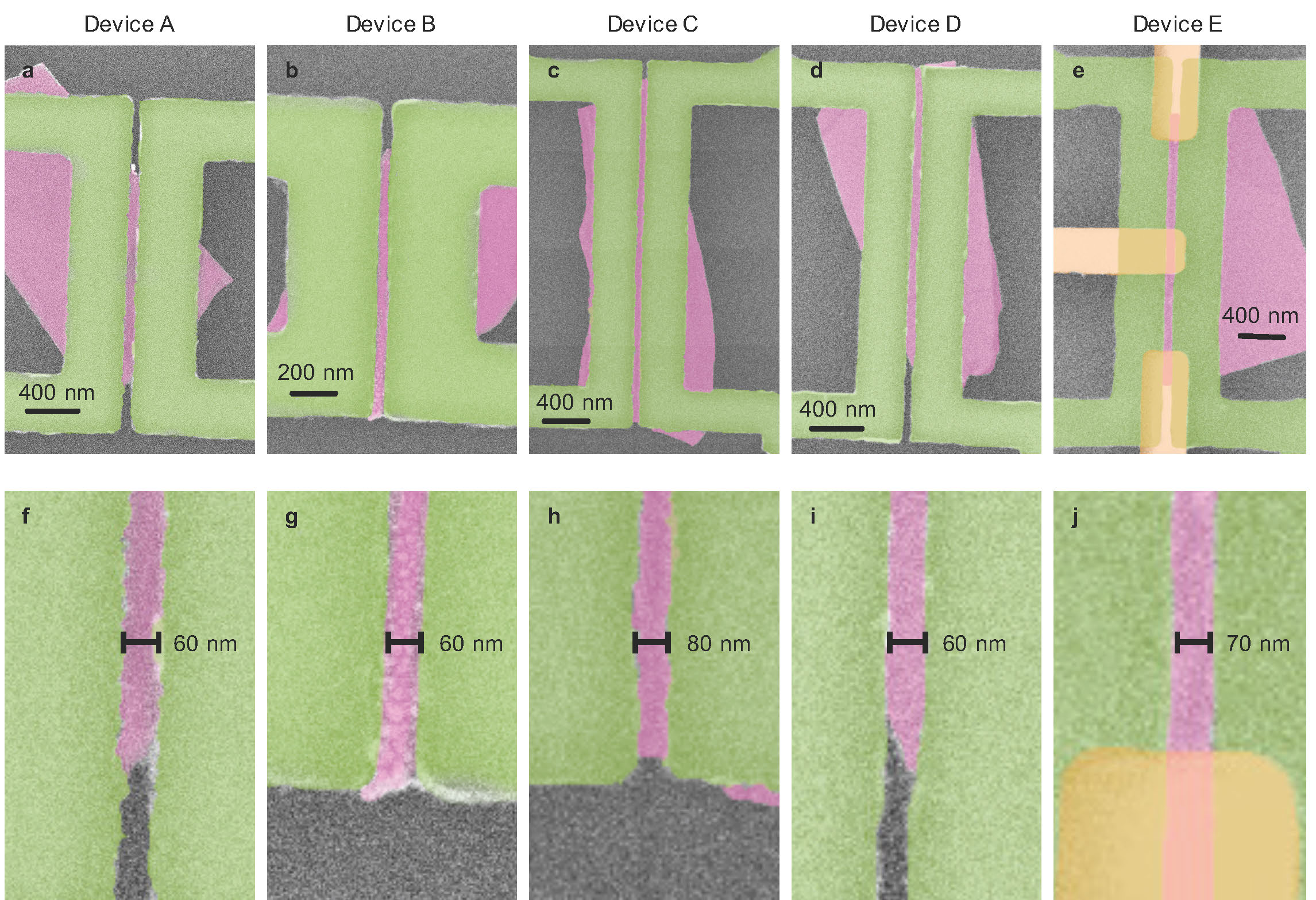}
	\caption{\textbf{SEM images of devices A--E.} (a-e) Global image. (f-j) Magnified view of the nanowire.}
	\label{fig:S1}
\end{figure}

After etching, the nanowires were contacted with superconducting electrodes from two sides along their full length by sputter-depositing Nb films of about 45\,nm thickness. The electrode width was about 400\,nm. In Fig. \ref{fig:S1}, we show the global SEM images of devices B and C (similar to that shown for device A in Fig. 1a of the main text) as well as the magnified views of the nanowires in devicea A--C.  The SiO$_2$ dielectric of device C was unfortunately broken and it was not gate-tunable, while devices A and B were gate-tunable. While a total of 5 devices were measured, we mainly report three of them, devices A-C. The other two, devices D and E, are used only to identify the relation between the nanowire cross-section and $I_c$-oscillation periodicity.

\vspace{-0.5cm}
\subsection{Transport measurements}
%\vspace{-0.5cm}

To electrically characterize the Josephson junctions, transport measurements were carried out at base temperature ($T \approx$ 30\,mK) of our dry dilution refrigerator (Oxford Instruments TRITON 300). For noise reduction, RC and copper-powder filters are installed in all electrical lines. In general, the differential resistance d$V$/d$I$ was measured in a quasi-four-terminal configuration; when this was not possible due to broken contact leads, we measured in a two-terminal configuration and subtracted the filter resistance. An AC current with a frequency below 40\,Hz and an amplitude of 5\,nA was superimposed to a DC bias current. A 6-1-1 T vector superconducting magnet enabled magnetotransport measurements with precise alignment of the magnetic field with respect to the nanowire axis. The temperature dependence of d$V$/d$I$ was measured in a second cool-down in a $^3$He refrigerator (Oxford Instruments Heliox VT) suitable for the desired temperature range.

%\newpage
%\vspace{0.8cm}
%{\large{\bf Supplementary Text}}
%\vspace{-0.5cm}

%\section{Supplemental Data and Discussions}

\vspace{-0.5cm}
\subsection{Temperature dependence of the critical current}
\vspace{-0.5cm}

\begin{figure}[h]
	\centering
	\includegraphics[width=0.4\textwidth]{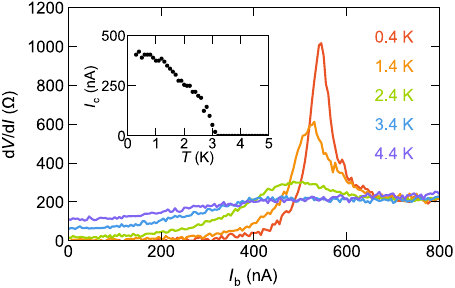}
	\caption{$dV/dI$ vs $I_{\rm b}$ data in device B at various $T$ for zero back-gate voltage; inset shows the $T$-dependence of the critical current $I_c$ extracted from these data. As in the main text, we define $I_c$ as the bias current above which $dI/dV$ exceeded 50 $\Omega$.}
	\label{fig:S2}
\end{figure}

The temperature dependence of the critical current $I_c$ in device B at zero back-gate voltage is shown in Fig. \ref{fig:S2} (similar data for device A was shown in Fig. 1c of the main text). The $I_c(T)$ curve is convex in both devices A and B, which suggests that our junctions are in the diffusive regime \cite{Dubos2001,Zhang2021}. 
%Since the diffusion constant and the Thouless energy are not well-characterized for BiSbTeSe$_2$ nanowires, we leave the quantitative analysis of the $I_c(T)$ behavior for future research. 

\vspace{-0.5cm}
\subsection{Junction transparency}
\vspace{-0.5cm}

\begin{figure}[h]
	\centering
	\includegraphics[width=0.4\textwidth]{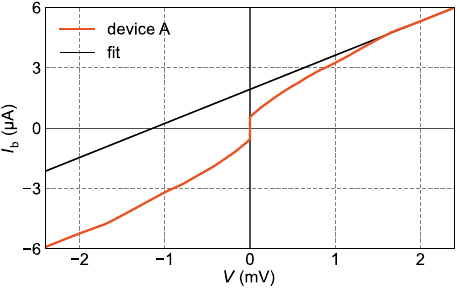}
	\caption{\textbf{$I$-$V$ characteristics.} $I$-$V$ characteristics measured in device A at 30\,mK in 0 T, together with a linear fit to the high-bias range with the slope constrained by $dV/dI$ = 590\,$\Omega$, which is obtained from the high-bias range of Fig. 1d of the main text.
	}
	\label{fig:S3}
\end{figure}

The $I$-$V$ curve of device A taken without any magnetic field nor gate voltage applied is plotted in Fig. \ref{fig:S3}. The critical current was $I_\mathrm{c}= 600$\,nA. The junction becomes resistive at higher bias currents, but linearity is only restored for $V_\mathrm{b}> 2\Delta_\mathrm{Nb}/e$ due to the multiple Andreev reflection (MAR) processes. Fitting a straight line to the high-bias linear regime with the slope constrained by the saturated $dV/dI$ value of 590\,$\Omega$ seen in Fig. 1d of the main text (which corresponds to the normal-state resistance $R_\mathrm{n}$), we obtain the excess current $I_\mathrm{e}$ = 1.93\,\textmu A from the intercept at $V_\mathrm{b}=0$. Based on the $R_\mathrm{n}$ and $I_\mathrm{e}$ values obtained this way, we estimate the barrier strength $Z$ = 0.56 which corresponds to the junction transparency $T_\mathrm{J} = (1+Z^2)^{-1} \approx 0.76$ in the Blonder-Tinkham-Klapwijk (BTK) theory \cite{Blonder1982, Octavio1983, Flensberg1988}.

\vspace{-0.5cm}
\subsection{Flux-focusing effect in out-of-plane and in-plane magnetic fields}
%\vspace{-0.5cm}

The flux expulsion due to the superconductor's diamagnetism increases the magnetic flux inside the junction. For out-of-plane fields, it is often accounted for by increasing the effective junction area to $L\cdot (W+W_\mathrm{SC}^{\rm left}/2+W_\mathrm{SC}^{\rm right}/2)$ with $W_\mathrm{SC}^{\rm left/right}$ the width of the superconducting electrodes \cite{Ghatak2018}. In our devices, $W_\mathrm{SC}^{\rm left} = W_\mathrm{SC}^{\rm right}$ = 400 nm. Note that in our sandwich TINW junctions, the usual terminology ``junction length'' corresponds to the nanowire width $W$ and  ``junction width'' corresponds to the nanowire length $L$. Based on this assumption, expected period $\Delta B_\mathrm{z}^\mathrm{eff}$ and observed periods $\Delta B_\mathrm{z}^\mathrm{exp}$ are in reasonable agreement, as summarized in Table \ref{table:Tab2}.

\addtolength{\tabcolsep}{5pt}    
\begin{table}[h]
\small
\centering
\begin{tabular}{ c c c c c}
 Device & $\Delta B_z^\mathrm{th}$ (mT) & $\Delta B_z^\mathrm{exp}$ (mT)\\ 
 \hline
  A & 5.3 & 5.8 \\ 
  B & 6.6 & 6.1 \\
    C & 3.1 & 2.2\\ 
\end{tabular}
\caption{Geometrically expected periods calculated from the effective areas considering the flux-focusing effect and experimentally observed magnetic-field periods for the out-of-plane ($z$-axis) magnetic fields for devices A--C.}
\label{table:Tab2}
\end{table}
\normalsize

\begin{figure}[h]
	\centering
\includegraphics[width=0.8\textwidth]{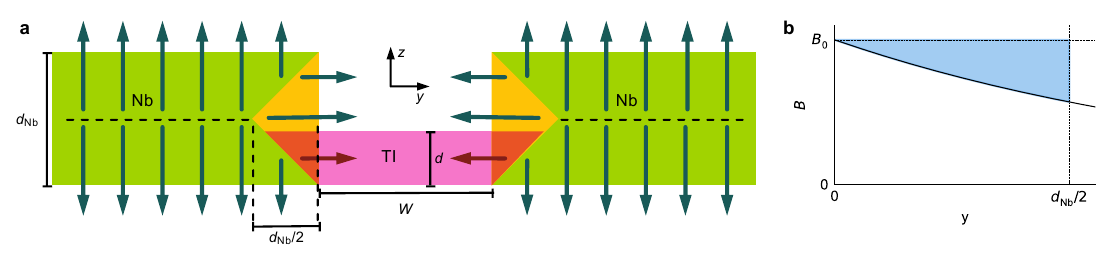}
	\caption{\textbf{Schematics of the in-plane flux-focusing effect.} (a) Schematic cross-sectional drawing of the Josephson junction indicating to which side magnetic field is expelled from the superconductor in the presence of an in-plane magnetic field. For flux focusing, only the red triangles need to be considered. (b) Magnetic-field profile $B(y)=B_0e^{-y/\lambda_\mathrm{Nb}}$ with $y$ the depth into the superconductor. The shaded blue region $B_0-B(y)$ indicates how much field is expelled.}
	\label{fig:S4}
\end{figure}

To consider the possible impact of the flux-focusing effect in the in-plane field geometry, we followed the model \cite{Molenaar2014, Assouline2019} to assume that a SC slab expels the magnetic fields in the manner shown by arrows in Fig.  \ref{fig:S4}a, either vertically or horizontally. When the thickness of the SC slab is smaller than or comparable to the London penetration depth, the spatial distribution of the magnetic-field penetration $B(y)=B_0e^{-y/\lambda_\mathrm{Nb}}$ (plotted in Fig. \ref{fig:S4}b, $B_0$ is the external magnetic field) should be taken into account. Here, we consider $\lambda_\mathrm{Nb}=40$\,nm, which is the London penetration depth of bulk Nb \cite{Maxfield1965}. Furthermore, since our TINW occupies only a part of the space between the SC electrodes, only the magnetic-field expulsion from the area shaded with red color in Fig.  \ref{fig:S4}a contribute to the enhancement of the magnetic flux in the TINW.
Based on these assumptions, the additional flux in the TINW due to the flux-focusing effect is calculated as
\begin{align*}
    \Phi_\mathrm{add} = 2\int_0^d\mathrm{d}z\int_0^z\mathrm{d}y\left(B_0-B(y)\right)=2B_0\bigg(\frac{d^2}{2}-\lambda_\mathrm{Nb}d+\lambda_\mathrm{Nb}^2\big(1-e^{-d/\lambda_\mathrm{Nb}}\big)\bigg)\,.
\end{align*} 
In terms of effectively added area, this results in $A_\mathrm{add}=\Phi_\mathrm{add}/B_0$. Even for the thickest TINW with $d=22$\,nm, this in-plane flux focusing only changes the expected period $\Delta B_\mathrm{x}$ by 4.4\,\%.

\vspace{-0.5cm}
\subsection{Irregular Fraunhofer patterns due to current-density inhomogeneity}
%\vspace{-0.5cm}

\begin{figure*}[h]
\centering
\includegraphics[width=0.8\textwidth] {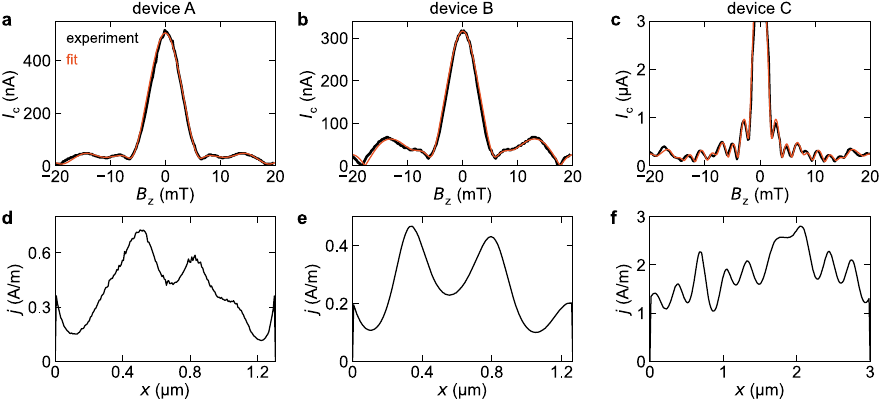}
\caption{{\bf Maximum-entropy fitting of the irregular Fraunhofer patterns and current-density-distribution retrieval.} (a-c) Data (black) and fitting (red) of the $B_z$-dependencies of $I_c$ in devices A--C in $B_x$ = 0 T. For devices A and B, the plotted $I_c$ is the average of $|I_c|$ for $+I$ and $-I$ directions to cancel the self-field effect. %For device C, since $I_c$ values were measured only for $+I$ direction, the plotted $I_c$ is the average of the data for $+B_z$ and $-B_z$, which cancels the asymmetry.
For device C, since $I_c$ was measured only for $+I$ direction, no such averaging was performed and the data contain the weak asymmetry that is presumably due to self-field effect \cite{Zhang2024}.
(d-f) The current-density distribution $j(x)$ obtained as a result of the maximum-entropy fitting \cite{Ghatak2018} to the data in panels (a-c); these forms of $j(x)$ give the red curves in (a-c).}

\label{fig:FraunhoferFits}
\end{figure*}

The Fraunhofer-like oscillations of $I_c$ with $B_z$ shown in Figs. 1e$-$1g of the main text for devices A$-$C exhibit irregular patterns, deviating strongly from the regular Fraunhofer pattern. As already shown in Ref. \cite{Ghatak2018}, these deviations can be explained as a result of inhomogeneous current-density distributions in the junction along the $x$-direction. 
Such inhomogeneities arise due to fabrication-related irregularities, including edge roughness and local variations in the interface transparency. 
One may notice a slight asymmetry in the $I_c(B_z)$ behavior for $\pm B_z$ in Figs. 1e$-$1g of the main text, but this is most likely an additional self-field effect from the supercurrent \cite{Zhang2024}.
To demonstrate that the observed irregular patterns can basically be explained by a current-density inhomogeneity, we applied the fitting procedure based on the maximum entropy method described in Ref. \cite{Ghatak2018} to reconstruct the possible current-density distribution $j(x)$ from the measured $I_c(B_z)$ behavior  (neglecting the weak asymmetry for $\pm B_z$). In Fig. \ref{fig:FraunhoferFits} we show the calculated patterns based on the $j(x)$ profiles shown at the bottom, which are the results of the maximum-entropy fitting, along with the experimental data. While the extracted $j(x)$ profiles are not unique solutions, the reasonable agreement between the data and calculations support the interpretation that the irregular Fraunhofer patterns originate essentially from inhomogeneous current-density distributions.

%\newpage
%\vspace{-0.5cm}
\subsection{Magnetic-field alignment to eliminate spurious $B_z$ component}
%\vspace{-0.5cm}

\begin{figure*}[h]
\centering
\includegraphics[width=0.7\textwidth] {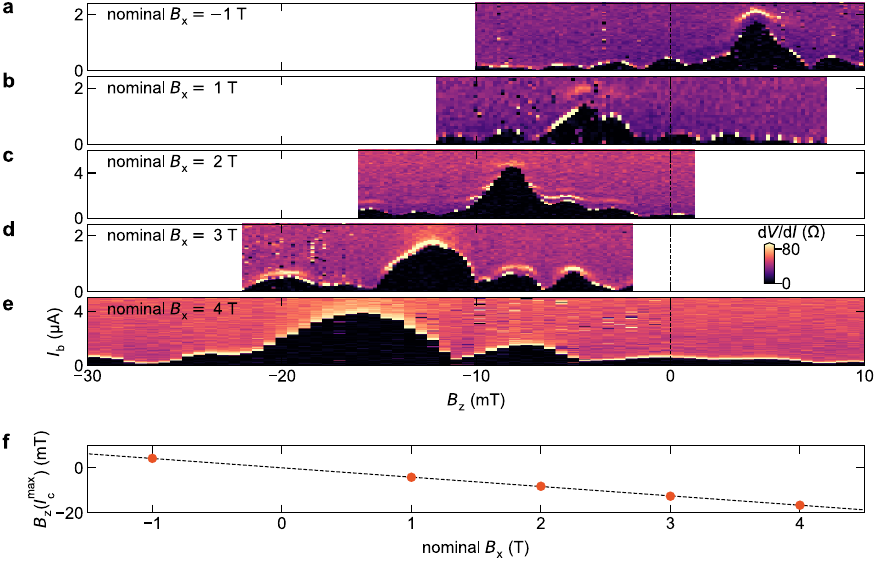}
\caption{{\bf Magnetic-field alignment procedure.} (a-e) $dV/dI$ as a function of $I_\mathrm{b}$ and $B_z$ measured in different $B_x$ fields at $T=30$\,mK in device C. The position of the maximum in the Fraunhofer-like pattern, $B_z(I_c^{\rm max})$, shifts away from $B_z=0$ due to a spurious $B_z$ component caused by the nominal $B_x$ field. (f) Plot of $B_z(I_c^{\rm max})$ vs nominal $B_x$ presenting a linear relation. This relation is used for determining the necessary compensation field $B_z(B_x)$ to achieve a pure in-plane field without any $B_z$ component.
The same alignment applies to all three devices that were fabricated to be parallel.}
\label{fig:alignment}
\end{figure*}

The alignment of the magnetic field with respect to the junction coordinate system is the key for our study of the critical current when the field is applied along the nanowire axis. If the alignment is imperfect, a nominal axial field $B_x$ can be accompanied by a finite $B_z$ component caused by misalignment, which creates a phase winding and a current modulation along the junction.  
To perform an accurate alignment, several critical-current measurements were done as a function of small $B_\mathrm{z}$ in the presence of a large nominal $B_x$. Figure \ref{fig:alignment} presents the observed Fraunhofer-like responses of device C in nominal $B_x = -1$, 1, 2, 3, and 4\,T; one can see that the position of the maximum critical current presents a systematic shift. 
As shown in Fig. \ref{fig:alignment}f, this shift is essentially linear in $B_\mathrm{x}$, implying that the shift can be attributed to a spurious $B_z$ component due to misalignment, which cancels the nominal $B_z$ to realize the actual $B_z$ = 0 at the critical-current maximum.  
This way, we can determine the necessary nominal $B_z$ to cancel the spurious $B_z$ component as a function $B_x$, such that we can obtain the $B_x$-dependence of the critical current in the absence of $B_z$. The observed slope of $-4.15$\,mT/T in Fig.~\ref{fig:alignment}f corresponds to a misalignment angle of only $\sim$0.2$^\circ$. Since devices A--C were all fabricated on the same wafer with NWs aligned in parallel, the same misalignment applies to all three devices.

Note that the maximum in the Fraunhofer pattern should appear at $B_z$ = 0 even in the presence of an anomalous phase $\varphi_0$ \cite{Buzdin2008} that is expected for a TI Josephson junction in a strong $B_x$-field \cite{Assouline2019}. This is because the global phase offset will self-tune to maximize the critical current in a current-biased Josephson junction. Nevertheless, the anomalous phase $\varphi_0$ leads to an asymmetry in the Fraunhofer pattern across $B_z$ = 0 \cite{Assouline2019}, which is actually apparent in our data in Figs. \ref{fig:alignment}a--e, suggesting that our sandwich TINW junction also develops an anomalous phase $\varphi_0$ in a large $B_x$. 
In passing, the increase in the period of the $B_z$ oscillations (i.e. Fraunhofer pattern) for increasing $B_x$ seen in Fig.~\ref{fig:alignment} is due to the change in the flux focusing effect in the junction; namely, a large $B_x$ field causes a significant fraction of the supercurrent density in the Nb electrodes to be used for shielding the $B_x$ field, making the flux expulsion from Nb for the $B_z$ field to become weaker and the effective area for the Fraunhofer pattern to become smaller.

\vspace{-0.5cm}
\subsection{Gate-voltage-dependence of the critical current in device A}
%\vspace{-0.5cm}

\begin{figure}[h]
	\centering
	\includegraphics[width=0.4\textwidth]{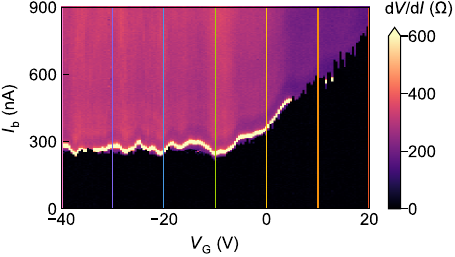}
	\caption{\textbf{$V_{\rm G}$-dependence of the critical current in device A.} 2D color mapping of d$V$/d$I$ measured in the $I_\mathrm{b}$-$V_\mathrm{G}$ plane in device A. Vertical lines indicate the positions at which the $B_x$ dependences of $I_\mathrm{c}$ shown in Figs. 2e of the main text were measured.}
	\label{fig:gating}
\end{figure}

In Fig. 2d of the main text, we showed $V_{\rm G}$-dependence of the critical current $I_c$ in device B.  Figure \ref{fig:gating} shows a similar plot for device A, where essentially the same behavior was observed. The saturation of $I_c$ in the $p$-type regime occurs for the same $V_{\rm G}$ range ($V_{\rm G} \lesssim -10$ V), and $I_c$ fluctuates in this range possibly due to the Fabry-Perot-like interference \cite{Calado2015, Ben2016}, although the fluctuations are more irregular than those in device B.

\vspace{-0.5cm}
\subsection{$I_c$ oscillations as a function of $B_x$ in devices D and E}
%\vspace{-0.5cm}

\begin{figure*}[h]
\centering
\includegraphics[width=0.72\textwidth]{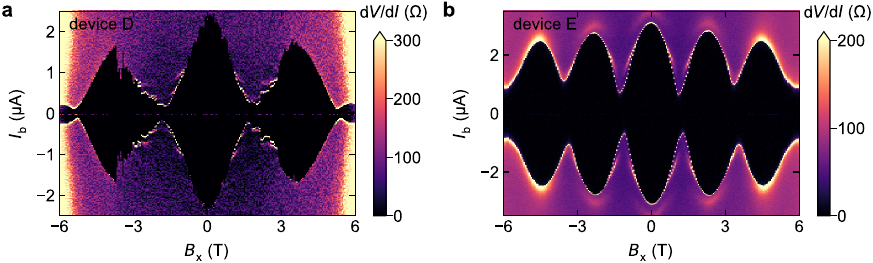}
\caption{{\bf SQUID-like $I_c$ oscillations in devices D and E.} (a-b) $dV/dI$ as a function of $I_\mathrm{b}$ and $B_\mathrm{x}$ obtained at $V_\mathrm{G}=0$\,V in devices D and E. The $B_z$-component is compensated to be zero. The sweep direction of $I_\mathrm{b}$ is different for positive and negative currents, such that the sweep always starts from 0 A. }
\label{fig:DE}
\end{figure*}

The $I_c$ oscillations as a function of $B_x$ in devices D and E are shown in Fig. \ref{fig:DE}. The periodicity $\Delta B_x$ shown in Fig. 3 of the main text for devices D and E were obtained from these data.

%\newpage
\vspace{-0.5cm}
\subsection{Gate-voltage dependence of the $B_x$-oscillation period}
%\vspace{-0.5cm}

\begin{figure*}[h]
\centering
\includegraphics[width=0.75\textwidth]{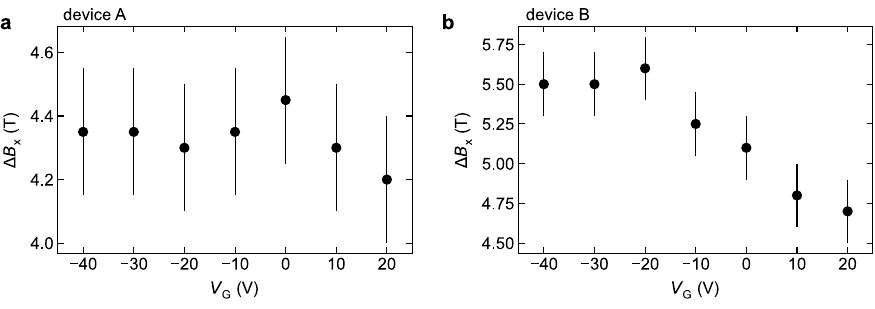}
\caption{{\bf $V_{\rm G}$ dependence of the SQUID-like oscillation period .} (a-b) $\Delta B_x$ as a function of $V_\mathrm{G}$ in devices A and B. Error bars reflect the uncertainty in identifying the $I_c$ minima.}
\label{fig:periodgating}
\end{figure*}

For both gate-tunable devices A and B, we present the $V_{\rm G}$ dependence of the SQUID-like oscillation period $\Delta B_x$ in Fig. \ref{fig:periodgating}. While $\Delta B_x$ remains largely unchanged in device A and remains within errors, $\Delta B_x$ in device B shows a non-negligible decrease towards more positive $V_\text{G}$. 
%We attribute this behavior to an effective increase of the nanowire area due to the positive electric field of the gate potential, attracting the charge carriers.
This decrease in $\Delta B_x$ is understandable if the positive gate voltage attracts more electrons towards to bottom surface, enlarging the electronic cross-section. 
However, the effect of gating on the distribution of the surface-electron wavefunction in a TI nanowire is sensitive to the chemical potential and the geometry \cite{Legg2021} and it can lead to both a slight increase and decrease in the effective cross-sectional area. We do not currently understand the difference between devices A and B.

%\newpage
\clearpage
%\vspace{-0.5cm}
\subsection{Transport simulations}
%\vspace{-0.5cm}

Here, we discuss further the transport simulations outlined in the main text and Methods section of the microscopic model of the TINW. Our simulations used the Kwant package \cite{Groth2014}.
In the normal state of a TINW, it is possible that the surface states are coherent along the full circumference. When this is the case, in the normal state, the eigenstate wavefunctions of a TINW can be indexed by quantised angular momenta, $\ell$, that takes half-integer values $\pm \frac{1}{2},\pm \frac{3}{2},\dots$ due to the Berry phase of the topological surface state~\cite{Zhang2010}. The corresponding quasi-one-dimensional subband dispersions are given by~\cite{Zhang2010,Cook2011,Legg2021}
\begin{equation}
E_{\ell}(k)=\pm\hbar v_\mathrm{F} \sqrt{k^2+\left(\frac{\ell- \Phi/\Phi_0^n}{R}\right)^2}, \label{eqn:Elk}
\end{equation}
where $v_\mathrm{F}$ is the Fermi velocity, $k$ is the momentum along the wire, $\Phi_0^n=2\Phi_0^s=h/e$ is the normal-state flux quantum. Here, a circular cross-section of radius $R$ is assumed, but this description is only weakly dependent on the cross-section shape \cite{deJuan2019}. It is important to notice that the normal-state band structure already depends on the flux $\Phi$, but with a periodicity of $h/e$. Although these subbands are doubly degenerate at $\Phi$ = 0, this degeneracy comes from the fact that the top and bottom surfaces of the TINW are taken together; if we look at the states locally, the top and bottom surfaces are individually described by the  BHZ model and are locally spin-non-degenerate. 
Our nano-SQUID model considers the junctions at the top and bottom surfaces locally. It should be noted that, due to the Dirac nature of the surface states, the Zeeman effect enters \eqref{eqn:Elk} in the same way as the orbital component of the magnetic field \cite{Legg2021}. As such, including the Zeeman effect would slightly renormalise the period of oscillations in a similar manner to how this period is modified by the extent of the surface state wavefunction into the bulk.
%, the double-degeneracy of the normal-state subband is irrelevant. 

We also note that for $\Phi = h/(2e)$, the lowest-energy subband has $\ell = \frac{1}{2}$ and becomes spin-non-degenerate. However, in the Bogoliubov-de Gennes Hamiltonian, its hole partner has $\ell = -\frac{1}{2}$ and hence the Cooper-pairing is prohibited by the orthogonality. This is known as the angular-momentum mismatch problem \cite{Cook2011, Cook2012, deJuan2019, Heffels2023}. This problem can be circumvented if the TINW is fully covered by a superconductor and a vortex is created \cite{Cook2012}.

\begin{figure}[h]
	\centering
	\includegraphics[width=0.82\textwidth]{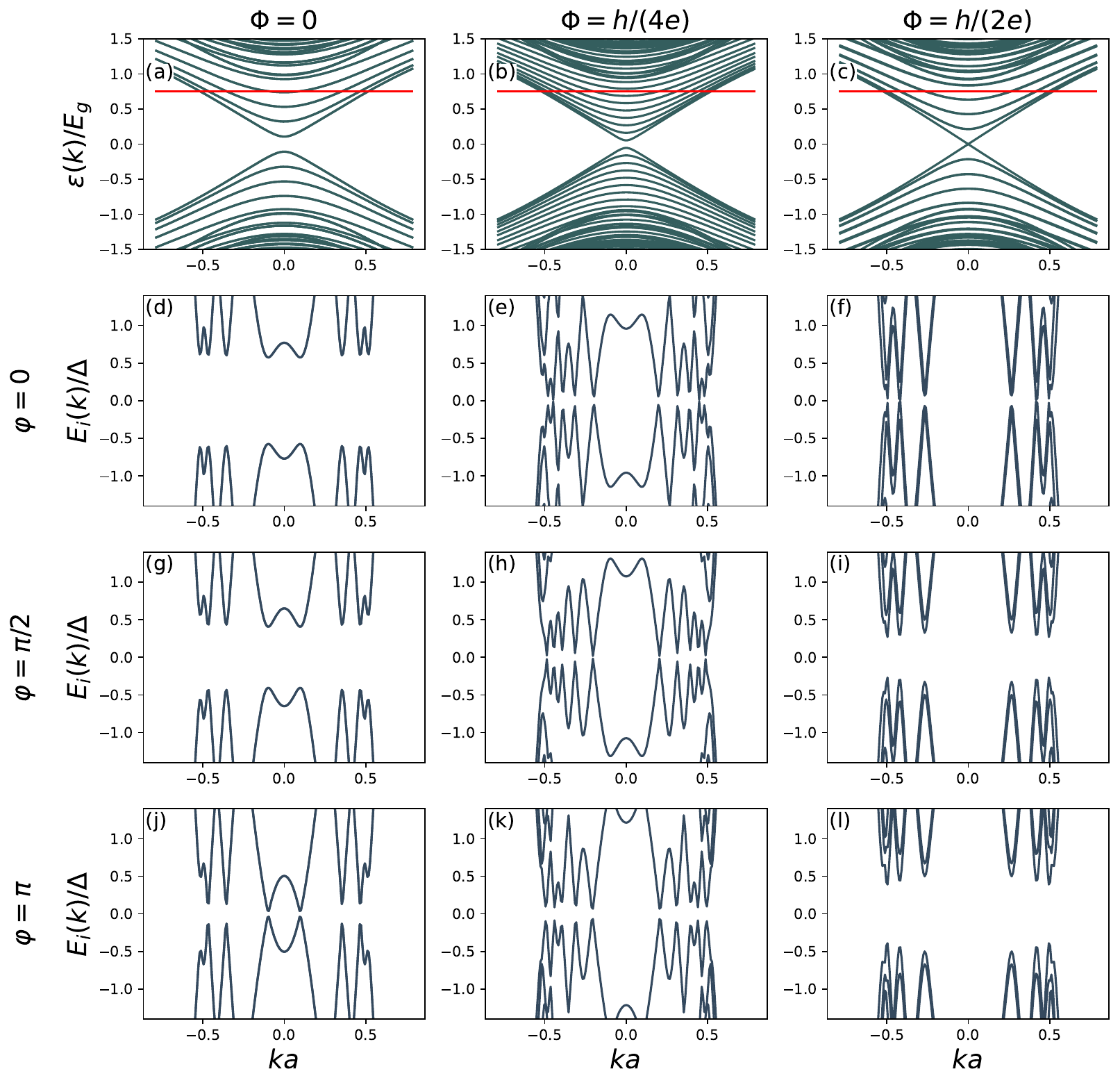}
	\caption{{\bf Normal-state and superconducting band structure of TINW nano-SQUID.} (a)-(c) The normal-state subband structure of the TINW  as a function of flux $\Phi$ calculated using the tight-binding model and parameters set out in Methods, but without any pairing potential. The bands at $\Phi=0$ are doubly degenerate. At finite flux this degeneracy is removed until $\Phi=h/(2e)$, where there is a gapless mode with $\ell = \frac{1}{2}$. The energy is normalized by the gap to the lowest bulk band, $E_g$, which for our parameters occurs at $\mu=0.663$. The chemical potential $\mu=0.5$ (red line) are set for all subsequent plots. (d-l) The BdG band structure of the TINW nano-SQUID with phase differences $\phi=0,\pi/2,$ and $\pi$ as indicated on the left. For $\Phi=0$ (d,g,j), the largest gap occurs for phase difference $\varphi=0$ indicating the ground state (minimum of energy phase relation, see Fig.~\ref{fig:phaserels}) occurs for this value. 
%The smallest gap occurs at $\phi=\pi$ indicating that the energy phase relation will have a maximum there. 
For flux $\Phi=h/(4e)$ (e,h,k), the size of the gap is only weakly dependent on $\varphi$, giving rise to a weak dependence of the energy phase relation on $\varphi$ in Fig.~\ref{fig:phaserels}). For flux $\Phi=h/(2e)$ (f,i,l), the largest gap occurs at $\varphi=\pi$ and the smallest at $\varphi=0$, indicating that the ground state now occurs at $\varphi=\pi$. 
%Since the gap for $\varphi=\pi$ is similar in size to the maximum that occurs when $\Phi=0$ the critical currents are approximately the same for both values of flux. 
The presence of a large gap at $\varphi=\pi$ for $\Phi=h/(2e)$ can be viewed as the superconductors providing an effective vortex.}
	\label{fig:bands}
\end{figure}

\begin{figure}[h]
	\centering
	\includegraphics[width=0.8\textwidth]{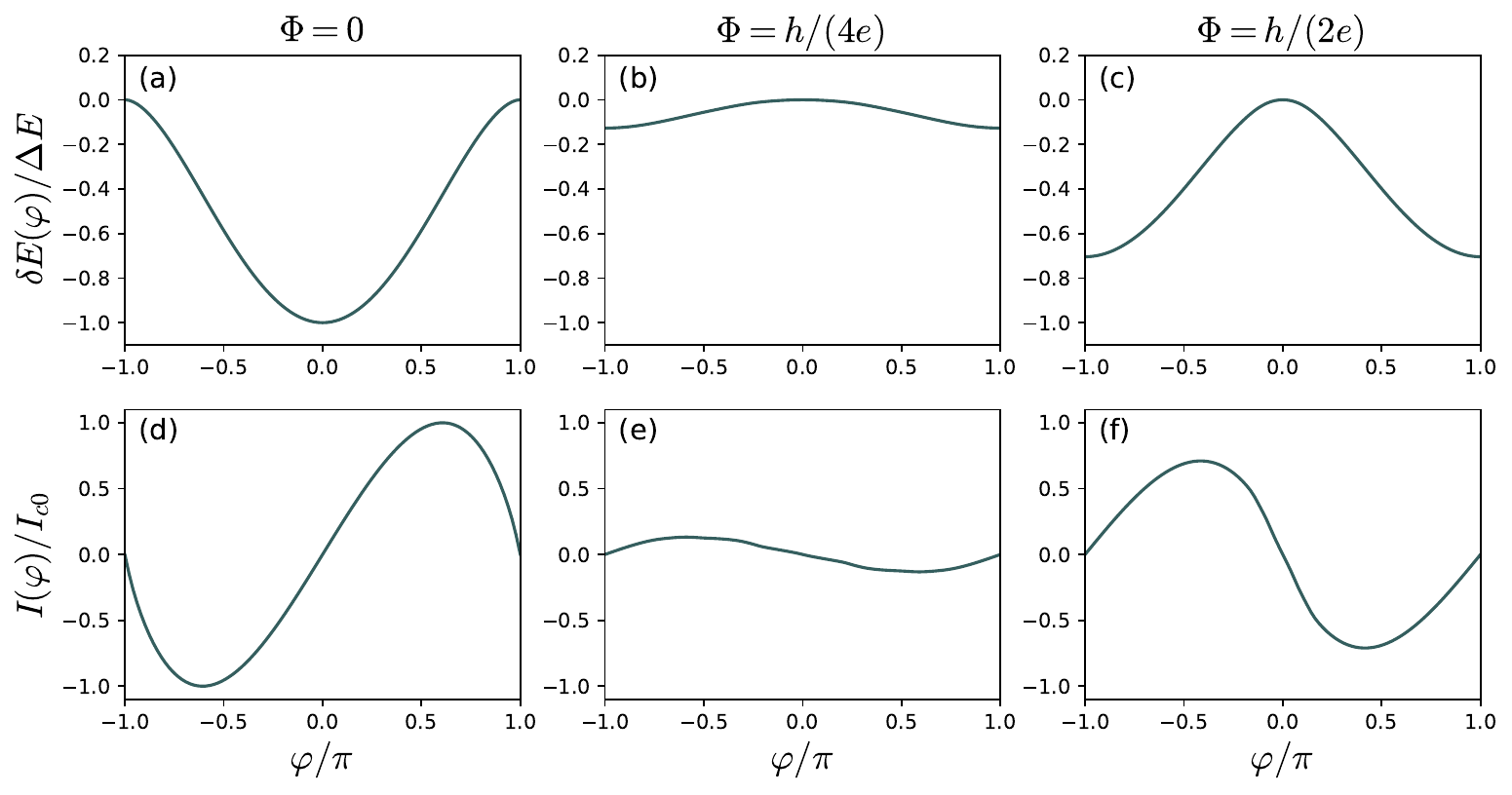}
	\caption{{\bf Energy and current phase relations.} (a)-(c) Energy change $\delta E(\varphi)=E(\varphi)-{\rm max}_\varphi\{E(\varphi)\}$ as a function of phase $\varphi$, normalize by the total change in energy $\Delta E={\rm max}_\phi\{E(\varphi)\}-{\rm min}_\phi\{E(\varphi)\}$ at $\Phi=0$. As expected, for $\Phi=0$ the ground state occurs at $\varphi=0$. In contrast, the ground state occurs at  $\varphi=\pi$ for $\Phi=h/(2e)$. The energy-phase relation is nearly independent of $\varphi$ for $\Phi=h/(4e)$. (d)-(f) Current-phase relations (CPRs) for different values of $\Phi$, normalized by the critical current at zero flux $I_{c}^{\Phi=0}$. The CPRs are the derivatives of the energy shown in panels (a)-(c). Whilst (d) and (f) have large maximum values, corresponding to large $I_c$ for $\Phi=0$ and $\Phi=h/(2e)$, the weak dependence of energy on phase for $\Phi=h/(4e)$ results in a very small $I_c$ as observed in experiment.}
	\label{fig:phaserels}
\end{figure}

The $I_c$ oscillations observed in our experiment can be understood from the SC band structures shown in Fig.~\ref{fig:bands}. In particular, for flux $\Phi=0$, the band structure has the largest gap for the phase difference $\varphi=0$ and the smallest for $\varphi=\pi$; this means that the ground state occurs at $\varphi=0$, because it is energetically favourable to realize a larger gap. In contrast, for $\Phi=h/(2e)$, the maximal gap occurs for phase difference $\varphi=\pi$ and so the ground state occurs when the system has a $\pi$ phase difference. This can be viewed as the phase difference providing an effective vortex that cancels the angular momentum mismatch of electrons and holes that occurs for this value of flux. This cancellation enables a gap to open that is similar in size to the gap which occurs at $\Phi=0$, resulting in a similar sized critical current. Finally, close to $\Phi=h/(4e)$, the SC band structure is essentially independent of phase and, correspondingly, the energy phase relation has only a very weak dependence on the phase difference $\varphi$ and the critical current is negligibly small. The energy-phase relations are illustrated in Fig.~\ref{fig:phaserels} for different fluxes.

\begin{figure}[h]
	\centering
	\includegraphics[width=0.75\textwidth]{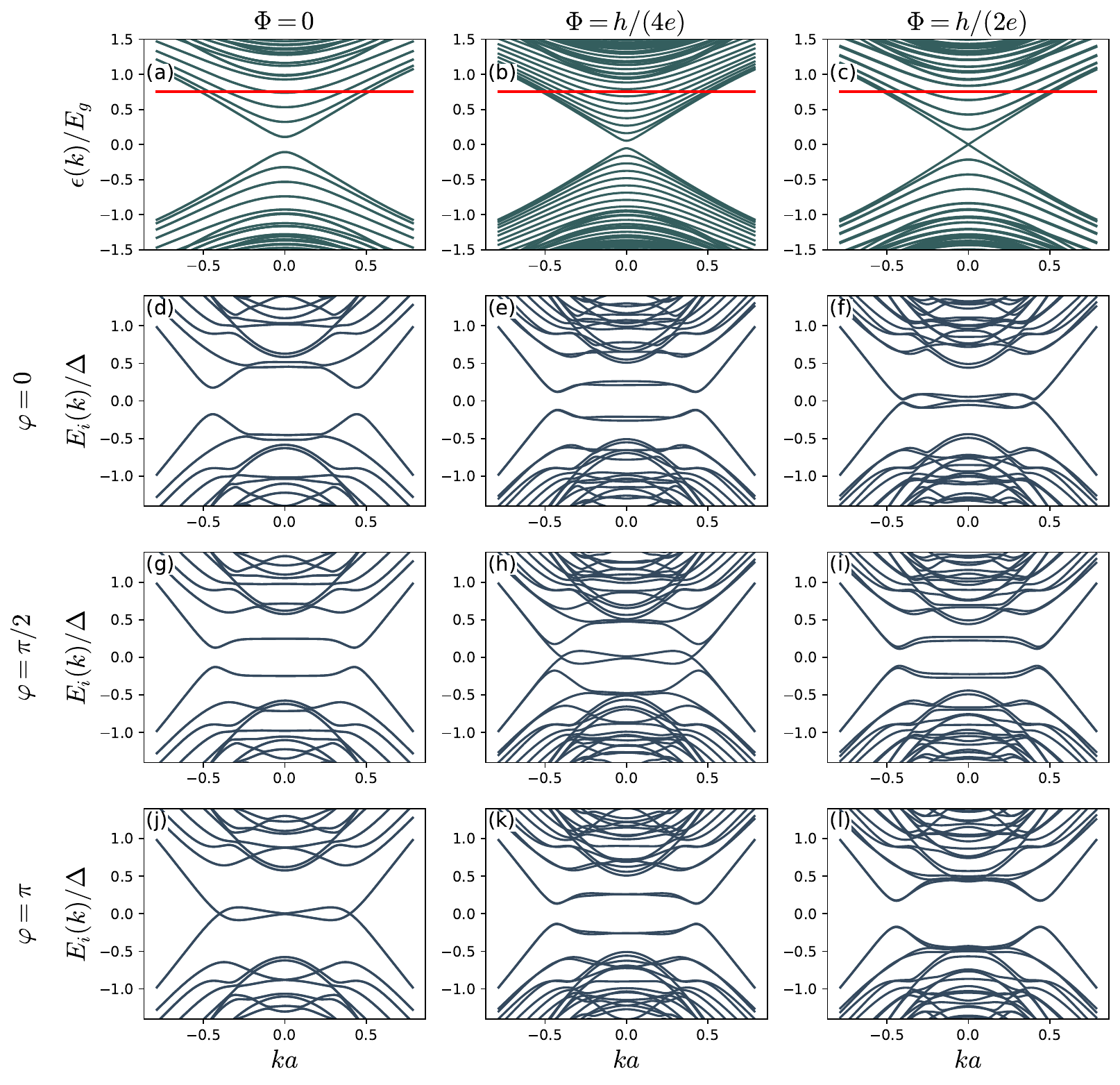}
   \vspace{-3mm}
   \caption{{\bf Normal-state and superconducting band structure of TINW nano-SQUID with large SC pairing.} Same as Fig.~\ref{fig:bands}, but for a pairing larger than the subband pacing (here $\Delta=0.25$). In such a setup the top and bottom surfaces are effectively decoupled, however due to the system forming a nano-SQUID, the energies are still periodic in $h/e$. Although our measured induced pairing is smaller than the subband gap, in reality, the decoupling of top and bottom surfaces can occur due to a multitude of factors, such as disorder e.g. at the interface of the TINW with the superconductor.}
	\label{fig:bands2}
\end{figure}

\begin{figure}[h]
	\centering
	\includegraphics[width=0.75\textwidth]{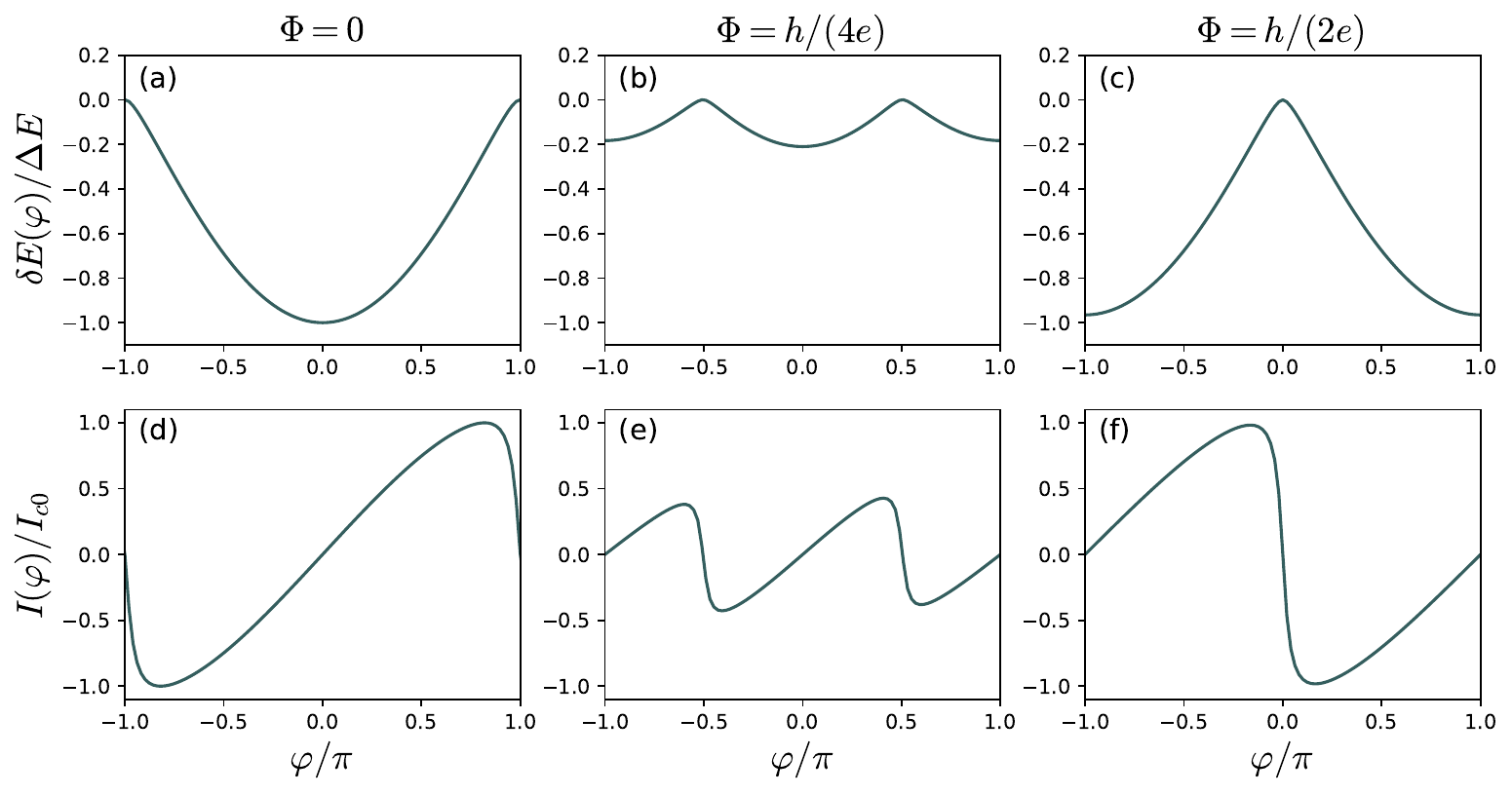}
    \vspace{-3mm}
	\caption{{\bf Energy and current phase relations for large SC pairing.} Same as Fig.~\ref{fig:phaserels}, but for a pairing larger than the subband spacing (here $\Delta=0.25 M$).  In such a setup the top and bottom surfaces are effectively decoupled, but the energies are still periodic in $h/e$ due to  nano-SQUID. Overall the critical current behaviour is essentially similar to that of the system with well-defined subbands. In particular, a $0-\pi$ phase transition occurs and the critical current is recovered $\Phi=h/(2e)$. As discussed in the main text, this scenario without well-defined subbands does not exhibit any even-odd effect.}
	\label{fig:phaserels2}
\end{figure}

We also consider the case where the SC pairing potential is larger than the subband gap, see Figs.~\ref{fig:bands2} \& \ref{fig:phaserels2}. This scenario is supposed to capture the situation where top and bottom surface are decoupled and subbands are no longer well defined. This decoupling can occur for several reasons, such as disorder at the interface between the TINW and the superconductor. Overall the critical current behaviour is similar to that of the system with well-defined subbands. Although we note that, due to the effective high-transparency of this setup, the energy phase relation close to $h/(4e)$ has two well developed minima that results in a $\pi$-periodic CPR close to this point i.e. $
\sim \sin(2\phi)$, see Fig.~\ref{fig:phaserels2}. However, it is natural that disorder will reduce the transparency and it is an interesting question whether further investigations can observe this effect. As discussed in the main text, this scenario also does not exhibit any even-odd effect, as expected since subbands are no longer well-defined.

Finally, it is important to note the difference between the effect observed here and the Little-Parks effect. In our model of the TI nanowire, the pairing potential is induced by the Nb superconducting leads and is not intrinsic to the TI nanowire itself. In other words, the relevant Ginzburg-Landau physics for the pairing potential occurs in the leads and not in the TI nanowire itself. The corresponding absence of fluxoid quantization in the SQUID means the absence of Little-Parks-like effect. The crucial factor resulting in the oscillations observed in our model and in the experiment is the phase difference between the leads, i.e., the junction phase difference itself.

%\clearpage

%\newpage

\vspace{-3mm}

\newpage
%\vspace{-0.5cm}
\subsection{Connection between subband picture and nano-SQUID picture}
%\vspace{-0.5cm}

We now turn to the connections between the subband picture from microscopic simulations and the nano-SQUID picture of two independent JJs on top and bottom surface. Assuming equal contributions from the top and bottom JJs, the full energy-phase relation of the SQUID is
\begin{equation}
E(\varphi)\approx-\varepsilon \cos(\varphi-\pi \Phi/\Phi_0^s)-\varepsilon \cos(\varphi+\pi \Phi/\Phi_0^s ) = -2 \epsilon \cos \varphi \cos (\pi \Phi/\Phi_0^s).\label{nanoS}
\end{equation}
Turning to the subband picture, a given state with a momentum $k$ has a wavefunction that lies at the exterior of the TINW and has a pairing potential only at the sides. As such, each state by itself effectively forms a nano-SQUID. In particular, within the subband picture, the energy phase relation is found by intergrating over all states in every occupied band, see Methods. Since each state by itself is effectively a nano-SQUID, this means that the CPR found from our microscopic simulation can be viewed as a sum over the CPRs of individual nano-SQUIDs, as in \eqref{nanoS}, and we can write
\begin{equation}
E(\varphi)=\sum_{E_i(k)<0} \int_{-\Lambda}^\Lambda dk\;E_i(k)\approx - \sum_i \varepsilon_i(\Phi) \cos \varphi - \varepsilon_0,\label{subbandSQUID}
\end{equation}
where $\varepsilon_0$ is the total phase independent energy and $ \sum_i \varepsilon_i(\Phi) \approxprop \cos (\pi \Phi/\Phi_0^s)$ is the sum of all the states in each occupied band. 

The approximation that the integral can be approximated by a sum over nano-SQUIDs, as in \eqref{subbandSQUID}, reproduces well our results, in particular, (i) $ \sum_i \varepsilon_i(\Phi) >0$ for $\Phi=0$, (ii) $ \sum_i \varepsilon_i(\Phi) \approx 0$ at $\Phi=h/(4e)$, and (iii) $ \sum_i \varepsilon_i(\Phi) <0$ for $\Phi=h/(2e)$. However, it should be noted that there can be further dependencies of $\varepsilon_i(\Phi)$ that are not captured by the simple approximation $\varepsilon_i(\Phi) \approxprop \cos (\pi \Phi/\Phi_0^s)$ due to the fact that the subbands in the normal state depend on flux.

It should be noted that here we modeled each Josephson junction (top/bottom) by a simple CPR of the form $I(\varphi)=I_{t/b} \sin \varphi$, i.e. a single Josephson harmonic. This is sufficient for the purpose of capturing the main effect of the SQUID-type oscillations. However, as can be seen in Fig.~4a of the main text, the microscopic tight-binding model will contain higher harmonics, $I(\varphi)=\sum_n I_{t/b}^{(n)} \sin (n \varphi$) with $n\in\mathbb{N}$, which are due to a reasonably high transparency of the junction. Including these harmonics would capture the full Fourier series of the CPR that is produced by the microscopic model, but the main SQUID-like effect can already be very well understood by the lowest order harmonic.

Finally, we estimate the superconducting coherence length on the side surface to determine the coherent tunneling probability between the top and bottom surfaces. 
For a top surface
with superconductivity induced by the proximity effect of Nb, we estimate a coherence length of $\xi_{{\rm t}}\approx75\,{\rm nm}$ (see Methods). For the side surface, we expect a much
shorter coherence length $\xi_{{\rm s}}\ll\xi_{{\rm t}}$ due to the strong
disorder on the etched surface, as well as a reduced Fermi velocity
of the modes moving in a plane perpendicular to the plane of the material layers.
When $\xi_{{\rm s}}$ is much smaller than the nanowire height $h=15\ {\rm nm}$,
electrons from the top and bottom surfaces undergo Andreev reflection before they
can tunnel to the other surface. This estimate suggests that the nano-SQUID with decoupled Josephson junctions would be a better approximation of the experimentally-realized device.
%interpretation of our system; however, coherent tunneling events between the top and bottom surfaces may reveal relevant contributions in microscopic modeling.

\clearpage
%\vspace{-0.5cm}
\subsection{Stability of the topological phase}
%\vspace{-0.5cm}

\begin{figure}[h]
	\centering
	\includegraphics[width=0.85\textwidth]{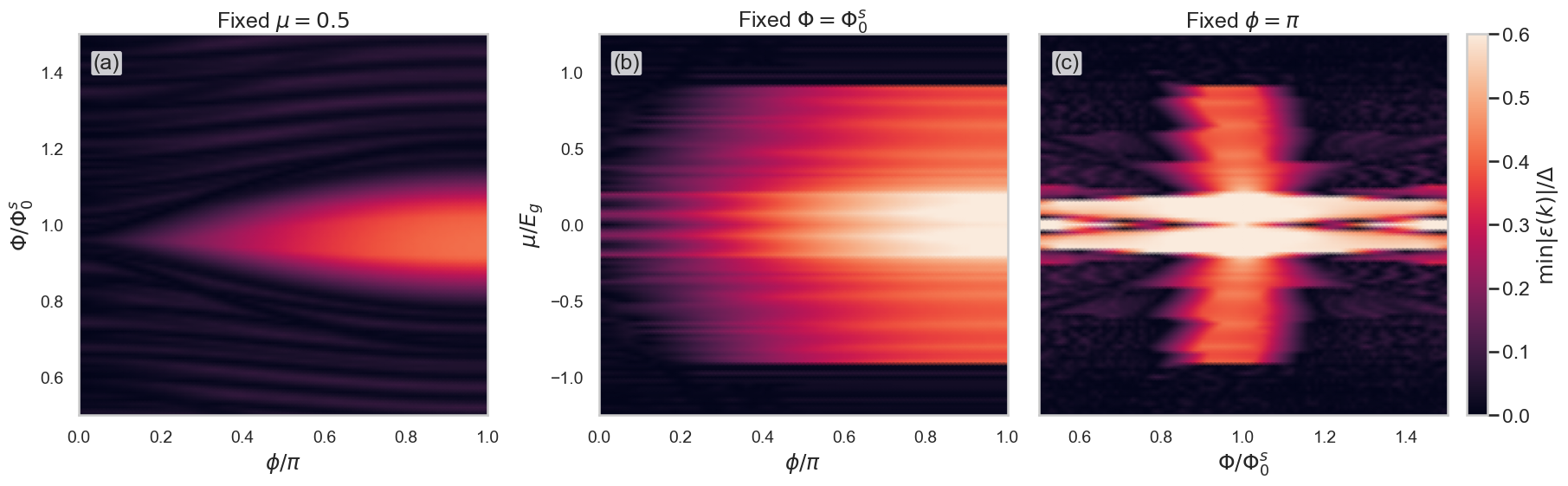}
	\caption{{\bf Stability of the topological phase with well-defined subbands.} There is a large region of phase space  containing a topological phase with sizable excitation gap, determined by the lowest energy as a function of momentum, $\min|\varepsilon(k)|$. (a) As a function of flux and phase difference across the junction for fixed chemical potential (red line in Fig.~\ref{fig:bands}) the optimal phase difference for the largest gap is $\phi=\pi$ and optimal flux is $\Phi=\Phi^s_0$, as would be expected from our nano-SQUID picture. (b-c) When the chemical potential is altered, we find that there is a gap to excitations that persists for the full bulk-band gap, $E_g$, indicating a very large stability to chemical potential fluctuations. Note that this bulk band gap is set by the lowest bulk state in the BHZ model. Finally, we note that there are finer features related to the presence of subbands in this simulations, e.g., the largest excitation gap occurs at the bottom of the first subband.}
	\label{fig:stability1}
\end{figure}

\begin{figure}[h]
	\centering
	\includegraphics[width=0.85\textwidth]{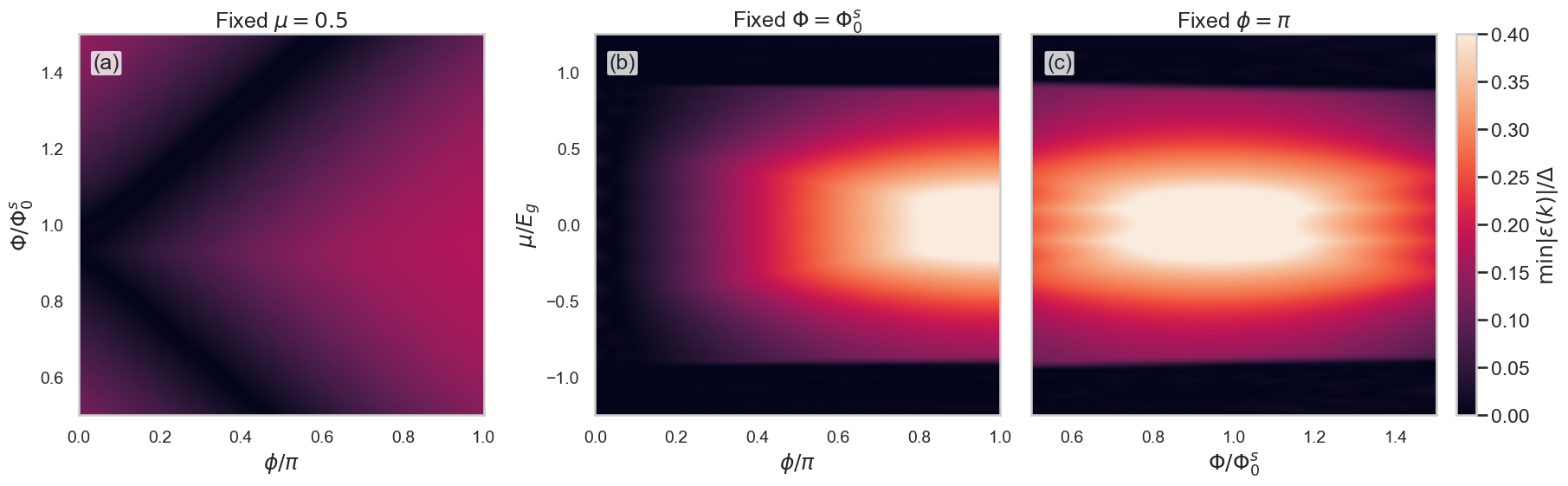}
\caption{{\bf Stability of the topological phase with decoupled Josephson junctions.} Similar to Fig.~\ref{fig:stability1}, but now with the parameters used for the decoupled Josephson junction simulations. The features are broader in this simulation due to the absence of well-defined subbands. Note that the optimal regime for the largest gap is essentially unchanged from the case of well-defined subbands. The broader feature means that the topological gap is less sensitive to small changes in the parameters.}
	\label{fig:stability2}
\end{figure}

We now analyse the stability of the topological phase to perturbations of parameters. In Figs.~\ref{fig:stability1}~\&~\ref{fig:stability2} the minimum energy to an excited state, $\min|\varepsilon(k)|$, is shown for the separate cases of well-defined subbands and decoupled junctions (as discussed above), using the same parameters as in our transport simulations. We find that in both scenarios there is a large region of phase space with a significant gap to excitations. In particular, within these simulations, a gapped topological phase is present for the full bulk band gap of the TI, $E_g$, demonstrating a very considerable stability to variations of chemical potential. Similarly, there is a very sizable range of both phase difference across the junction and flux through the nanowire where a gapped phase is maintained. As such, the large region of topological phase space strongly suggests that there should be a sizable tolerance to the inevitable disorder that will occur in these parameters.

Furthermore,  Figs.~\ref{fig:stability1}~\&~\ref{fig:stability2} provides a clear sweet spot for devices to be tuned into where topological superconductivity with a large superconducting gap and associated Majorana zero-modes should be present. In both scenarios the middle of the bulk band gap, close to the Dirac point, has the largest gap. It should also be noted that in the subband scenario there are finer features, such that the largest gap occurs close to the bottom of the first subband. Finally, in both cases, the largest gap to excitations occurs for a phase difference $\varphi=\pi$ and for a flux $\Phi=\Phi^s_0$, as would be expected from the nano-SQUID picture above. Between the two scenarios, the absence of finer structure makes it preferable to realize the decoupled junctions by inducing a large SC pairing potential in the TI.
%The fact that this regime is achieved in our experiments, evinces that topological insulator nano-SQUIDs are a highly promising platform for future experimental studies of topological superconductivity.}

%\clearpage
\newpage
\bibliography{nanoSQUID}